\documentclass[12pt]{iopart}

\usepackage{epsfig}

\def\asca       {{\em ASCA}\/}
\def\suzaku     {{\em Suzaku}\/}
\def\chandra    {{\em Chandra}\/}
\def\xmm        {{\em XMM-Newton}\/}
\def\rosat      {{\em ROSAT}\/}

\def\vla        {{\em VLA}\/}

\begin{document}
%\linenumbers

\title{Hot Gas in Galaxy Groups: Recent Observations}

\author{M Sun}

%\address{Department of Astronomy, University of Virginia, P.O. Box 400325, Charlottesville, VA 22904, USA}
\address{Eureka Scientific, Inc., 2452 Delmer Street, Suite 100, Oakland, CA 94602, USA}
\ead{mingsun.cluster@gmail.com}

\begin{abstract}

Galaxy groups are the least massive systems where the bulk of baryons begin to be accounted for.
Not simply the scaled-down versions of rich clusters following self-similar relations,
galaxy groups are ideal systems to study baryon physics, which is important
for both cluster cosmology and galaxy formation.
We review the recent observational results on the hot gas in
galaxy groups. The first part of the paper is on the scaling relations, including
X-ray luminosity, entropy, gas fraction, baryon fraction and metal abundance.
Compared to clusters, groups have a lower fraction of hot gas around the center (e.g., $r < r_{2500}$),
but may have a comparable gas fraction at large radii (e.g., $r_{2500} < r < r_{500}$).
Better constraints on the group gas and
baryon fractions require sample studies with different selection functions and deep
observations at $r > r_{500}$ regions. The hot gas in groups is also iron poor at
large radii (0.3 $r_{500}$ - 0.7 $r_{500}$). The iron content of the hot gas
within the central regions ($r <$ 0.3 $r_{500}$) correlates with the group mass,
in contrast to the trend of the stellar mass fraction. It remains to be seen where the
missing iron in low-mass groups is.
In the second part, we discuss several aspects of X-ray cool cores in galaxy groups,
including their difference from cluster cool cores, radio AGN heating in groups and
the cold gas in group cool cores.
Because of the vulnerability of the group cool cores to radio AGN heating and the
weak heat conduction in groups, group cool cores are important systems to test the
AGN feedback models and the multiphase cool core models.
At the end of the paper, some outstanding questions are listed.

\end{abstract}

\submitto{\NJP}
\maketitle

\section{Introduction}

Galaxy groups are less massive, gravitationally bound systems than galaxy clusters.
As in clusters, a significant fraction of baryons is in the hot, X-ray emitting gas
between galaxies.
The galaxy groups discussed in this paper are systems with the virial mass
between 10$^{12.5} h^{-1}$ M$_{\odot}$ and $10^{14.2} h^{-1}$ M$_{\odot}$.
\footnote{The virial mass ($M_{\rm vir}$) is the total mass within the virial radius.
The virial radius (or $r_{200}$) is defined as the radius at which the mean density of
the system is 200 times the critical density of the universe. Two common scale radii
in this paper are $r_{500}$ and $r_{2500}$, or the radii at which the mean density of
the system is 500 times and 2500 times the critical density of the universe, respectively.
The system mass referred in this paper is usually
$M_{500}$, the total mass contained within $r_{500}$. Typically, $r_{500} \approx$ 0.66 $r_{200}$,
$r_{2500} \approx$ 0.47 $r_{500}$
and $M_{500} \approx$ 1.4 $M_{200}$. The quoted mass range corresponds to gas
temperatures of $kT_{500} \approx$ 0.25 - 2.8 keV, where $T_{500}$ is
the temperature of the hot gas measured from the integrated spectrum in
the projected 0.15 $r_{500}$ - $r_{500}$ annulus.}
For a continuous halo mass function, why would we single out galaxy groups?
The short answer is that galaxy groups are not simply scaled-down versions
of rich clusters (e.g., Ponman et al. 1999; Mulchaey 2000;
Ponman et al. 2003; Voit 2005).
Because of shallow gravitational potential, galaxy groups are
systems where the roles of complex baryon physics (e.g., cooling,
galactic winds, and AGN feedback) become significant.
Galaxy groups are important for at least three reasons.
Firstly, groups are the least massive systems where most baryons begin to be
accounted for. Baryon physics can be studied for both the hot gas
(e.g., their thermal properties and the heavy element enrichment)
and the stellar component, as well as the energy and mass transfers between these
two components. The two baryon components also have different
mass dependence, with groups poorer in the hot gas content and richer in the stellar
fraction than clusters. Together, they provide crucial constraints on many
baryon physics that is essential to understand galaxy formation
(e.g., Ponman et al. 1999; Ponman et al. 2003; Bower et al. 2008).
Secondly, the effects of the same baryon physics are not large but still
significant in massive clusters, and therefore need to be calibrated if
we want to further improve the cosmological constraints from clusters
(e.g., Voit 2005; Vikhlinin et al. 2009, V09 hereafter).
Thirdly, there are many more groups than clusters and most galaxies in the local
universe are in groups.
Groups are also the building blocks of future clusters through hierarchical mergers.
Thus, galaxy group is a critical environment to study galaxy evolution.

This paper focuses on the recent observational results on the X-ray emitting gas in
galaxy groups. Our understanding of this vital baryon component in galaxy groups
has been significantly improved over the last decade, with the \chandra, \xmm\ and \suzaku\ data.
This paper is not meant to be a historic review, which can be found in e.g.,
Mulchaey et al. (2000), Ponman et al. (2003), Mulchaey et al. (2003) and Voit (2005).
The plan of this paper is as follows. Section 2 introduces the recent X-ray samples of
nearby groups and clusters discussed in this paper. Section 3 is on the X-ray luminosity
scaling relations of groups. The radial properties of the hot gas in groups are outlined
in Section 4. The baryon budget in groups is examined in Section 5.
Section 6 reviews the metal enrichment of the hot gas in groups. Several aspects of
cool cores and AGN heating in galaxy groups are discussed in Section 7.
Section 8 summarizes what has been discussed in this paper and presents an incomplete
list of outstanding questions for galaxy groups.
Two appendices are included at the end of the paper for the comparison on the X-ray luminosity
and the gas fraction results. 
All the work included in this paper is based on AtomDB 1.3.1 or previous versions.
\footnote{The AtomDB 2.0 released in 2011 includes significant changes on the iron L-shell
data, which affects the spectral fits for $kT < 2$ keV plasma. Temperatures increase
by 10\% - 20\%, while abundance decreases by $\sim$ 20\%, from AtomDB 1.3.1 to AtomDB 2.0.1.
However, works published prior to
2011 used AtomDB 1.3.1 or older versions, including the analysis of
the mock data from simulations. The AtomDB change needs to be kept in mind for
future comparisons.}
We assume that $\Omega$$_{\rm M}$=0.24, $\Omega_{\rm \Lambda}$=0.76 and
H$_{0}$ = 73 km s$^{-1}$ Mpc$^{-1}$.

\section{Recent X-ray sample studies of galaxy groups and clusters}

There are some recent works on X-ray samples of local galaxy groups, especially
with the \chandra\ and the \xmm\ data, which include:

\begin{itemize}

\item The \rosat\ sample of the Group Evolution Multiwavelength Study (GEMS) project
(Osmond \& Ponman 2004, OP04 hereafter): it is a study of 60 GEMS groups with the pointed
\rosat\ PSPC observations. We only include 33 groups in the G-sample (see OP04 for the detail)
with both temperature and luminosity measured. The inclusion of this sample increases the
temperature range of the X-ray luminosity scaling discussed in \S3.
The redshift range is 0.004 - 0.025, with a median of 0.013.
The group temperature ranges from 0.2 keV to 1.5 keV.

\item The two-dimensional \xmm\ Group Survey (2dXGS) sample (Finoguenov et al. 2006, 2007a)
and the Mahdavi et al. (2005) sample: both are \xmm\ archival samples analyzed in the
same way. The parent samples are X-ray flux limited samples.
These papers focus on two-dimensional properties of the hot gas, entropy and
pressure profiles. Johnson et al. (2009, 2011) combined these samples to study the entropy
distribution, the abundance profile and the gas core in groups.
The Johnson et al. sample has 28 groups at $z$ = 0.003 - 0.038, with a median of 0.014.
The range of the group temperature is 0.6 - 1.9 keV.

\item The \chandra\ sample by Vikhlinin et al. (2006, V06 hereafter) and V09: these two papers
produce an archival sample of 20 systems with the X-ray hydrostatic equilibrium (HSE) mass derived to
$r_{500}$ for 17 of them. There are six groups at $z$ = 0.015 - 0.081 and all of them are
also in the Sun et al. (2009) sample.

\item The Gastaldello et al. (2007, G07 hereafter) sample: it is an archival sample of 16 X-ray
bright groups at $z$ = 0.009 - 0.081, with a median of 0.024.
Seven of the 16 groups have both the \xmm\ and the \chandra\ data analyzed. Six groups only have
the \xmm\ data analyzed, while the other three groups only have the \chandra\ data analyzed.
For 13 groups, the gas properties at $r >$ 100 kpc (or $r > 0.25 - 0.6 r_{2500}$) were constrained
solely from the \xmm\ data. The group temperature ranges from 0.8 keV to 2.6 keV.
For the derivation of the X-ray HSE mass profile,
G07 used the temperature-based forward fitting method (see Buote \& Humphrey 2012 for detail)
that is different from the traditional approach used in most of the other works (e.g., V06 and Sun et al. 2009). 
G07 also examined the changes of their results with different mass modeling
approaches. 

\item The \chandra\ group sample by Rasmussen \& Ponman (2007, 2009): it is an archival sample
of 15 X-ray bright groups at $z$ = 0.005 - 0.027, with a median of 0.016. The parent sample is
the G-sample from OP04. The group temperature ranges from 0.3 keV to 2.1 keV.
This work is focused on the temperature profile, the abundance distribution and the implications
on feedback (see \S6).

\item The \chandra\ group sample by Sun et al. (2009, S09 hereafter): it is an archival sample
of 43 groups, all with pointed \chandra\ observations. Seventeen groups also have pointed
\rosat\ PSPC observations that were analyzed for the surface brightness profile at the group
outskirts. The redshift range is 0.012 - 0.122, with a median of 0.033.
$kT_{500}$ = 0.7 - 2.7 keV.

\item The Eckmiller et al. (2011, E11 hereafter) sample: it consists of 26 groups at $z$ = 0.012 - 0.049,
with a median of 0.024. The parent sample is an X-ray flux-limited sample of 112 groups
and the E11 subsample is \chandra\ archive limited.
The temperature range is 0.6 - 3.0 keV.
The main results are the gas mass and the total mass at $r_{2500}$ and $r_{500}$.

\end{itemize}

%%%%%%%%%%%%%%%%%%%%%%%%%%%%%%%%%%%%%%%%%%%%%%%%%
\begin{figure}[t]
\centerline{\includegraphics[width=0.8\textwidth,angle=270]{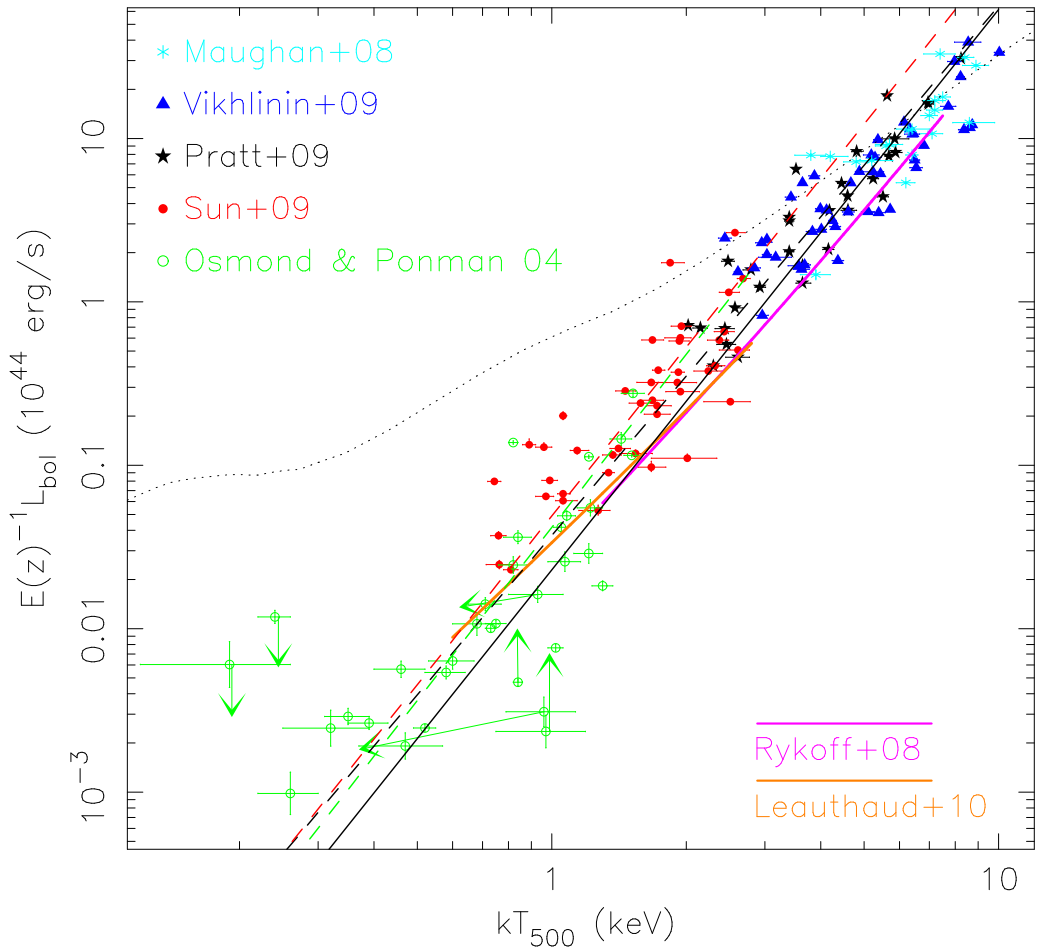}}
\caption{The bolometric luminosity (within $R_{500}$) - temperature
relation for $z < 0.2$ groups and clusters, with the data from OP04 (\rosat),
Maughan et al. (2008) (\chandra), S09 (\chandra), V09 (\chandra) and P09 (\xmm).
The red, black and green dashed lines are the best-fit relations for S09, P09 and OP04, respectively,
with the  BCES orthogonal regression method (Akritas \& Bershady 1996).
The black solid line is the best fit from P09 after the Malmquist bias is corrected
($\propto T_{500}^{3.43}$). The $M_{\rm HSE} - T$ relation from S09 (consistent with the
$M_{\rm HSE} - T$
relation from Arnaud et al. 2005) was used to convert the $L - M_{\rm HSE}$ relation from P09 to the
$L - T$ relation shown. The dotted line, $L_{\rm bol} \propto T^{1.5} \Lambda(T)$ that is under the
assumption of constant gas fraction and structure function, approaches the self-similar
relation ($L_{\rm bol} \propto T^{2}$)
for clusters and flattens to $L_{\rm bol} \propto T^{\sim 1.3}$ for groups.
An abundance of 0.3 solar is assumed. The normalization
of the dotted line is adjusted to match hot clusters.
While the \rosat\ results are included for more temperature coverage, the uncertainties
are larger than those from \chandra\ and \xmm\ (see \S3 and Appendix A for detail).
As a full analysis of the OP04 sample with the \chandra\ or the \xmm\ data is not available,
the trend of changes for six outliers in OP04 are shown by arrows (see Appendix A for detail).
The $L - M$ relations from MAXBCG (Rykoff et al. 2008b; Rozo et al. 2009) and COSMOS
(Leauthaud et al. 2010) are also shown by the magenta line and the orange line,
respectively.
}
\end{figure}
%%%%%%%%%%%%%%%%%%%%%%%%%%%%%%%%%%%%%%%%%%%%%%%%%

It is useful to compare the results from these sample studies, as often the same observations
were analyzed and the same scaling relations were examined. The comparison of the X-ray
luminosity scaling, between OP04 and S09, is presented in appendix A.
The comparison of the abundance results is in \S6. 
One of the key questions on galaxy groups is the baryon budget (see \S5).
The results on the gas fraction and total mass from V06, G07, S09 and E11
are discussed in appendix B.
The results from V06, G07 and S09 agree within 10\% - 20\%, while the E11
results are off to a larger extent.

Often in this paper, the cluster results are required for more mass coverage of the
scaling relations. The following cluster samples are mainly used.

\begin{itemize}

\item The \chandra\ cluster sample by Maughan et al. (2008): it is an archival sample
of 115 clusters at $z$ = 0.11 - 1.3. We only include 20 clusters at $z$ = 0.113 - 0.199
for the studies of X-ray luminosity and cool core. $kT_{500}$ = 3.8 - 8.9 keV.

\item The \chandra\ cluster sample by V09: the full sample (Table 2 of V09) is an X-ray flux
limited sample. We include 47 clusters at $z$ = 0.029 - 0.151 for the studies of X-ray
luminosity and cool core. $kT_{500}$ = 2.4 - 10.0 keV.
In combination with the V06 work, V09 also has a sample of 14 clusters with the X-ray HSE mass
derived to $r_{500}$.

\item The Representative \xmm\ Cluster Structure Survey (REXCESS) sample (B\"ohringer et al. 2007;
Pratt et al. 2009, P09 hereafter; Pratt et al. 2010):
it is an X-ray luminosity selected sample. The P09 work includes 31 clusters at $z$ = 0.056 - 0.183.
$kT_{500}$ = 2.0 - 8.2 keV.

\end{itemize}

\section{X-ray luminosity of the hot gas in galaxy groups}

The X-ray luminosity of the hot gas in groups and clusters
can be written as (Arnaud \& Evrard 1999):
$E(z)^{-1} L = f_{\rm gas}^{2}(T) [E(z) M(T)] \Lambda(T, Z) \hat{Q}$($T$),
where $f_{\rm gas}$ is the gas fraction and $M$ is the total mass.
$\Lambda(T, Z)$ is the cooling function that mainly depends on $T$ for
$kT > 2$ keV gas, but becomes more abundance ($Z$) dependent at $kT < 2$ keV.
In the above form, $\Lambda(T, Z)$ should be considered as an average cooling function as
there are temperature and abundance gradients in the hot gas.
$E^{2}(z)$ = $\Omega$$_{\rm M} (1+z)^{3} + \Omega_{\rm \Lambda}$.
$\hat{Q}$($T$) is the structure function that is equal to
$< \rho_{\rm gas}^{2} > / < \rho_{\rm gas} >^{2}$, averaging over the
cluster atmosphere.
\footnote{This is a three-dimensional definition, while the observed luminosity
is projected. In this paper, $r$ is the three-dimensional radius while $R$
is the projected radius. The difference between the three-dimensional
luminosity within $r_{500}$ and the projected luminosity within $R_{500}$ depends on
the density profile at $r > r_{500}$ and is usually small.
The projected luminosity can be up to 15\%
larger than the three-dimensional luminosity for groups, but the difference is
generally only $\sim$ 5\% for clusters.}
Thus, as one of the easiest properties of the hot gas to measure, the X-ray luminosity
carries important information on the gas distribution.

There have been a number of works on the scaling relations of X-ray luminosity (especially the
$L - T$ relation, e.g., Helsdon \& Ponman 2000; OP04; Maughan et al. 2008; P09),
although not many updates with the \chandra\ and \xmm\ results have been published so far.
In Fig.~1, we plot the $L_{\rm bol} (R < R_{500}) - T_{500}$ relation for 174 nearby
(or $z < 0.2$) groups and clusters. The group and cluster samples are introduced
in \S2.
The detail on the derivation of X-ray luminosities for the S09 groups is given
in appendix A, as well as the comparison with the \rosat\ results of OP04.
Fig.~1 includes the \rosat\ sample from OP04 for more temperature coverage.
However, the limitation of the PSPC data makes the OP04 results
suffer from larger uncertainties than the other results in Fig.~1, which is also discussed
in appendix A.
For clusters, both the \chandra\ and the \xmm\ samples are included, but extra caution
is required for the systematic difference of temperatures
between \chandra\ and \xmm\ (especially for hot clusters, e.g., Nevalainen et al. 2010).

The best-fit relations from the S09 group sample, the REXCESS sample and the OP04 G-sample
agree well (Fig.~1). The S09 sample has a larger intrinsic scatter than the REXCESS sample,
0.710$\pm$0.077 vs. 0.471$\pm$0.033 respectively ($\sigma_{\rm ln, intrinsic}$ in Table 1).
The best-fit relations from the S09 sample and the REXCESS sample are listed in Table 1,
as well as the scatter.
The Malmquist-bias corrected relation for the REXCESS clusters (P09) is also
shown in Fig.~1, while such a correction cannot be made for the S09 groups.
Such a correction steepens the $L - T$ relation.
The good agreement between S09 and the REXCESS results (P09)
also extends to the $L - Y_{\rm X}$ and the $L - M_{\rm HSE}$ relations.
Fig.~1 also includes the converted $L - T$ relations from the $L - M$ relations from the
MAXBCG clusters (Rykoff et al. 2008b; Rozo et al. 2009; Rozo, private communication) and
the COSMOS groups (Leauthaud et al. 2010). The $M_{\rm HSE,~500} - T_{500}$ relation from
S09 was used in the conversion.
MAXBCG is an optically selected cluster sample and the mass is derived from the
stacked lensing data. The COSMOS groups are X-ray selected (Finoguenov et al. 2007b)
and the mass is also derived from the stacked lensing data. The X-ray bolometric correction
has been done for both relations as the X-ray luminosities quoted in these papers are
not bolometric values. As shown in Fig.~1, the best-fit relations from MAXBCG and COSMOS
works compare well with the relations from the work with only the X-ray analysis. The difference between
the MAXBCG relation and the REXCESS relation at high mass can be explained by $\sim$ 20\%
mass bias (or $M_{\rm HSE,~500} / M_{\rm true,~500} \sim$  0.8). While the COSMOS results from
Leauthaud et al. (2010) may imply a small mass bias for groups, the current uncertainties
are still large.

We conclude that while groups are significantly X-ray fainter than the extrapolation
from clusters assuming constant gas fraction and structure function,
there is no unambiguous evidence for the steepening of the $L - T$ relation
at $kT <$ 2 keV (also see OP04), although the relation should inevitably steepen
at low mass (e.g., for single galaxies).
The scatter of the $L - T$ relation increases for groups.
While there are more results on the low temperature (or mass) part of the $L - T$
(or $L - M$) relation, better constraints requires group samples with well-defined
selection functions and better understanding of the bias for the X-ray HSE mass.

%%%%%%%%%%%%%%%%%%%%%%%%%%%%%%%%%%%%%%%%%%%%%%%%%
\begin{figure}[t]
\centerline{\includegraphics[width=0.7\textwidth,angle=270]{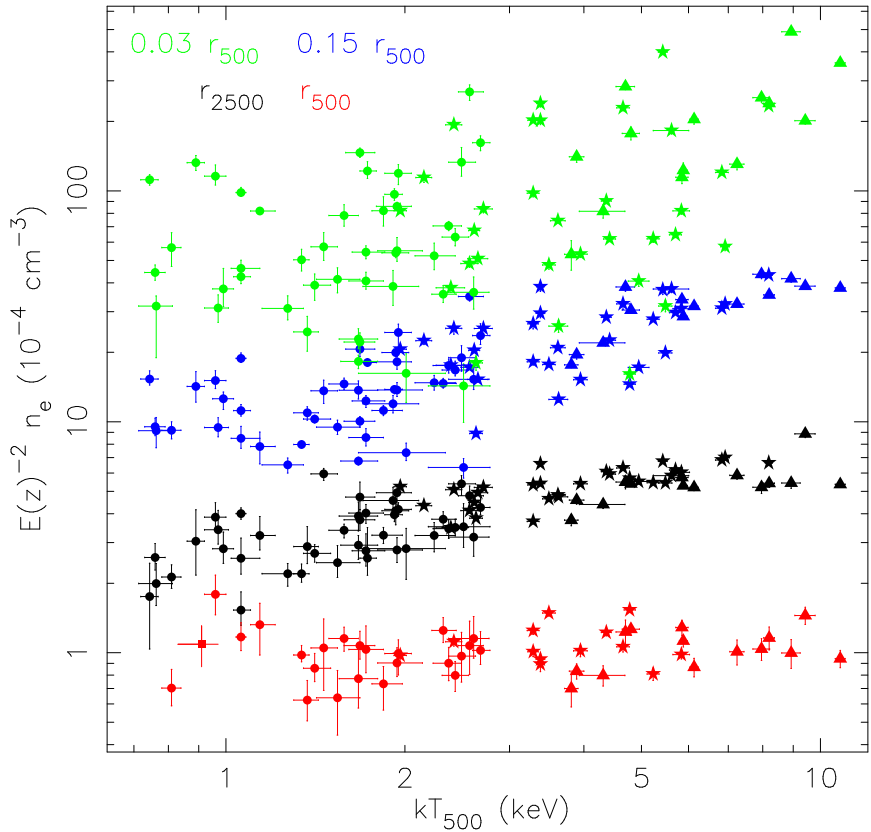}}
\caption{Electron densities at several characteristic radii for groups and
clusters from S09 (circles), P09 (stars), V09 (triangles) and Rasmussen et al. (2010)
(square, one point). The best fits are listed in Table 1 (including the relation
at $r_{1000}$). While groups have lower densities than clusters at inner radii,
the electron density at $r_{500}$ is almost independent of mass.
}
\end{figure}
%%%%%%%%%%%%%%%%%%%%%%%%%%%%%%%%%%%%%%%%%%%%%%%%%

\section{Radial properties of the hot gas in galaxy groups}

We can learn more about the thermodynamics of the hot gas from the radial distribution.
Groups generally have decreasing temperature profiles at $r >$ 0.2 $r_{500}$, similar
to clusters (e.g., V06; G07; Pratt et al. 2007; Rasmussen \& Ponman 2007; S09; Johnson et al. 2009).
On average, temperature profiles of groups have a prominent peak at 0.1 - 0.2 $r_{500}$,
while cluster temperature profiles are less peaky with a plateau at 0.15 - 0.3 $r_{500}$
(e.g., Sun et al. 2003; V06; Rasmussen \& Ponma 2007; Leccardi \& Molendi 2008a; S09).
Fig.~2 shows the electron density of the gas at several characteristic radii,
0.03 $r_{500}$, 0.15 $r_{500}$, $r_{2500}$ ($\sim$ 0.45 $r_{500}$),
$r_{1000}$ ($\sim$ 0.73 $r_{500}$) and $r_{500}$. The best fits and scatter are listed
in Table 1. The gas density is less mass dependent with increasing radius, even if the group data
are ignored. The scatter of the density scaling relation also becomes smaller at $r_{2500}$
and beyond, while the scatter at 0.03 $r_{500}$ is large. At $r_{500}$, the density is
almost a constant at $\sim 10^{-4}$ cm$^{-3}$. For the classical self-similar form (or both
the gas fraction and structure function are mass independent), gas densities at scaled radii
are mass independent.
Thus, while groups have less gas at the inner regions (e.g., $r < r_{2500}$) than clusters
to deviate the luminosity scaling relation from the self-similar form, the gas density relation
approaches the self-similar behavior at $r_{500}$.

Groups have elevated entropy profiles within $r_{500}$
(Ponman et al. 1999; Lloyd-Davies et al. 2000; Ponman et al. 2003;
S09; Johnson et al. 2009; Pratt et al. 2010).
The best fits and the scatter of the entropy - temperature relations at several characteristic
radii are listed in Table 1. The entropy - temperature relations of the groups
from S09 agree well with those from the REXCESS cluster sample
(Pratt et al. 2010).
Just as the behavior of the density scaling relations, the entropy - temperature
relations have smaller scatter at $r > r_{2500}$ and the slope is gradually approaching
the self-similar value. At $r_{500}$, there is no significant entropy excess above the
entropy baseline (Voit et al. 2005) , at least from the current samples that are mainly X-ray selected
(S09; Johnson et al. 2009; Pratt et al. 2010).
As the trend is to approach the slope of the self-similar scaling at large radii,
groups have entropy profiles that are flatter ($K \propto r^{0.6-0.8}$) than the baseline
entropy profile of $K \propto r^{1.1}$ (e.g., S09; Johnson et al. 2009).

%%%%%%%%%%%%%%%%%%%%%%%%%%%%%%
\begin{table}
\begin{center}
\caption{Scaling relations of luminosity, density and entropy \\}
\begin{tabular}{cccc} \hline \hline
Relation$^{\rm a}$, samples$^{\rm b}$ & $A$ & $\alpha$ & $\sigma_{\rm ln, intrinsic}$ ($\sigma_{\rm ln, data}$)$^{\rm c}$ \\ \hline

$L_{1}$ - $T_{500}$, S+R & 0.880$\pm$0.060 & 3.03$\pm$0.11 & 0.562$\pm$0.023 (0.164) \\
$L_{2}$ - $T_{500}$, S+R & 0.506$\pm$0.022 & 2.90$\pm$0.07 & 0.365$\pm$0.014 (0.161) \\
$L_{3}$ - $T_{500}$, S+R & 0.333$\pm$0.024 & 2.74$\pm$0.12 & 0.586$\pm$0.027 (0.148) \\
$L_{4}$ - $T_{500}$, S+R & 0.190$\pm$0.008 & 2.60$\pm$0.08 & 0.369$\pm$0.014 (0.144) \\
$L_{1}$ - $M_{\rm HSE,~500}$, S & 0.476$\pm$0.069 & 2.04$\pm$0.33 & 0.586$\pm$0.118 (0.457) \\
$L_{3}$ - $M_{\rm HSE,~500}$, S & 0.243$\pm$0.017 & 1.67$\pm$0.19 & 0.632$\pm$0.133 (0.439) \\
$n_{\rm e}$ (0.03 $r_{500}$) - $T_{500}$, S+R+V & 72.9$\pm$5.5 & 0.554$\pm$0.110 & 0.715$\pm$0.095 (0.097) \\
$n_{\rm e}$ (0.15 $r_{500}$) - $T_{500}$, S+R+V & 17.3$\pm$0.7 & 0.703$\pm$0.056 & 0.328$\pm$0.011 (0.072) \\
$n_{\rm e}$ ($r_{2500}$) - $T_{500}$, S+R+V & 4.05$\pm$0.10 & 0.448$\pm$0.039 & 0.182$\pm$0.006 (0.115) \\
$n_{\rm e}$ ($r_{1000}$) - $T_{500}$, S+R+V & 1.86$\pm$0.06 & 0.241$\pm$0.058 & 0.179$\pm$0.010 (0.130) \\
$n_{\rm e}$ ($r_{500}$) - $T_{500}$, S+R+V & 1.00$\pm$0.04 & 0.063$\pm$0.060 & 0.175$\pm$0.007 (0.152) \\
$K$ ($r_{2500}$) - $T_{500}$, S+R+V & 488$\pm$8 & 0.731$\pm$0.024 & 0.103$\pm$0.003 (0.124) \\
$K$ ($r_{1000}$) - $T_{500}$, S+R+V & 670$\pm$13 & 0.861$\pm$0.032 & 0.070$\pm$0.006 (0.161) \\
$K$ ($r_{500}$) - $T_{500}$, S+R+V & 812$\pm$28 & 0.952$\pm$0.053 & 0.125$\pm$0.007 (0.268) \\

\hline \hline
\end{tabular}
\item[] $^{\rm a}$
$E(z)^{-1} L = L_{0}$ ($T_{500}$ / 2.5 keV)$^{\alpha}$,
$E(z)^{-7/3} L_{\rm bol} = L_{0}$ ($M_{\rm HSE,~500}$ / 6$\times10^{13}$ h$^{-1}$ M$_{\odot}$)$^{\alpha}$,
$E(z)^{-2} L_{\rm 0.5 - 2\ keV} = L_{0}$ ($M_{\rm HSE,~500}$ / 6$\times10^{13}$ h$^{-1}$ M$_{\odot}$)$^{\alpha}$,
$E(z)^{-2} n_{\rm e} = n_{e, 0}$ ($T_{500}$ / 2.5 keV)$^{\alpha}$ and
$E(z)^{4/3} K = K_{0}$ ($T_{500}$ / 2.5 keV)$^{\alpha}$.
$L_{1}$ = $L_{\rm bol}$ ($< R_{500})$, $L_{2}$ = $L_{\rm bol}$~(0.15~$R_{500} - R_{500})$,
$L_{3}$ = $L_{\rm 0.5 - 2\ keV}$ ($< R_{500})$, $L_{4}$ = $L_{\rm 0.5 - 2\ keV}$ (0.15 $R_{500} - R_{500})$.
$A$ is either $L_{0}$ (10$^{44}$ erg s$^{-1}$), or $n_{e, 0}$ (10$^{-4}$ cm$^{-3}$)
or $K_{0}$ (keV cm$^{2}$).
The BCES orthogonal regression method (Akritas \& Bershady 1996) was used, except for
the density relation at 0.03 $R_{500}$ where the scatter is the largest and BCES(Y$\mid$X) was used.
\item[] $^{\rm b}$
``S'' = the S09 sample, ``R'' = the REXCESS sample and ``V'' = the V09 sample.
\item[] $^{\rm c}$
The intrinsic scatter and the statistical uncertainty of the data in brackets are shown.
\end{center}
\end{table}
%%%%%%%%%%%%%%%%%%%%%%%%%%%%%%

%%%%%%%%%%%%%%%%%%%%%%%%%%%%%%%%%%%%%%%%%%%%%%%%%
\begin{figure}[t]
\centerline{\includegraphics[width=0.62\textwidth,angle=270]{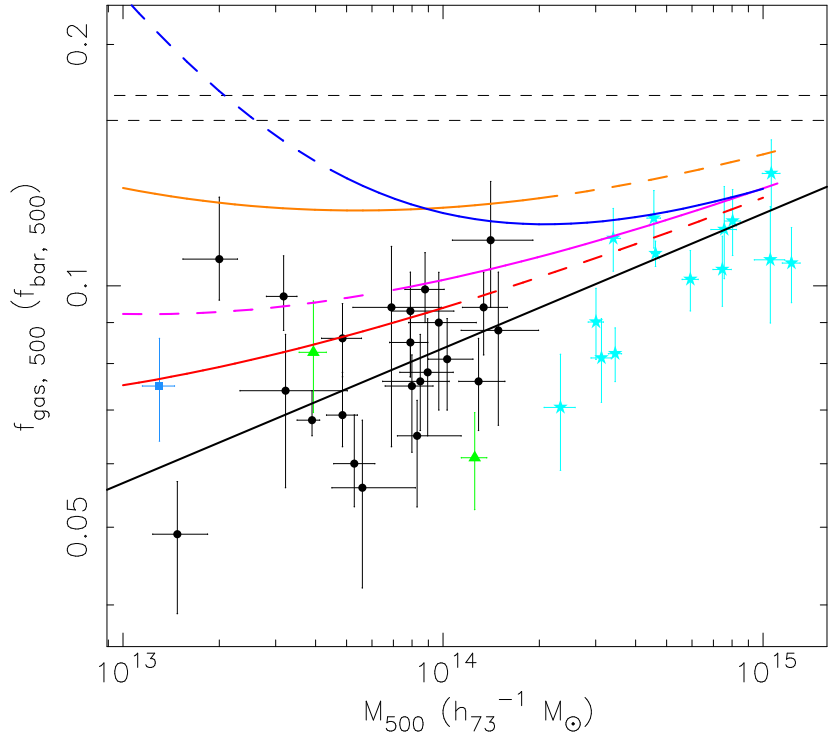}}
\caption{Gas and baryon fractions of groups and clusters.
The black, cyan, green and lightblue data points are from S09, V09,
D\'emocl\`es et al. (2010) and Rasmussen et al. (2010) respectively, while the
black solid line is the best fit.
The magenta, blue, orange and red lines are the expected baryon fractions
after adding the best-fit gas fraction (black line) with the stellar
mass fraction from Lin et al. (2003), Gonzalez et al. (2007), Giodini et al. (2009)
and Leauthaud et al. (2012) respectively (see Leauthaud et al. 2012 for the discussion
on the stellar mass fraction). Solid lines show the covered mass ranges from those works,
while dashed lines are extrapolations. 
Gonzalez et al. (2007) included the intracluster light, which is not included in other works.
The two black dashed lines enclose the universal baryon fraction
(0.1668$\pm$0.0060, Komatsu et al. 2011).
}
\end{figure}
%%%%%%%%%%%%%%%%%%%%%%%%%%%%%%%%%%%%%%%%%%%%%%%%%

While the pressure information is embedded in the relations for density and entropy,
it is discussed here for the connection to the signals from the Sunyaev–--Zel'dovich (SZ) effect.
The thermal SZ power spectrum began to be accurately measured by {\em SPT} and {\em ACT}
(Lueker et al. 2010; Shirokoff et al. 2011; Dunkley et al. 2011; Reichardt et al. 2011),
which opens a new window into high-redshift groups.
Half of the SZ power at $\ell$ = 3000 comes from halos with $10^{13} h^{-1}$
M$_{\odot} < M_{500} < 1.5\times10^{14} h^{-1}$ M$_{\odot}$ at $z > 0.5$
(e.g., Shaw et al. 2010; Trac et al. 2011). 
The measurement of the thermal SZ power spectrum shows a deficiency
of the SZ signals relative to the model (Sehgal et al. 2010; Shirokoff et al. 2011;
Dunkley et al. 2011).
Arnaud et al. (2010) used the function form initially suggested by Nagai et al. (2007)
to derive the ``universal pressure profile'' from the \xmm\ results on the REXCESS clusters.
Sun et al. (2011) shows that the group pressure profiles from S09 agree
with this ``universal pressure profile''.
The thermal SZ power spectrum scales roughly as the square of the thermal SZ flux.
Roughly the SZ signal,
$Y_{\rm SZ} \propto (M_{\rm gas} / M_{\rm HSE}) (M_{\rm HSE} / M_{\rm true})^{1.61} M_{\rm true}^{1.61}$,
while $M_{\rm true}$ is the true mass and $M_{\rm HSE}$ is the X-ray HSE mass.
The first factor is the gas fraction from the X-ray measurement, while the second factor is
the mass bias factor. We assumed $M_{\rm HSE} \propto T_{\rm X}^{1.65}$ from S09.
Any non-thermal pressure in the hot gas (e.g., turbulence) will result in the incomplete
thermalization of the potential energy, which makes the HSE mass measured from the X-ray data
biased low.
As shown by Shaw et al. (2010) and Trac et al. (2011), $\sim$ 20\% non-thermal pressure, being mass
independent, can reconcile the difference between observations and models on the SZ power spectrum
at $\ell$ = 3000 to $\sim$ 1 $\sigma$. Other potential solutions have also been discussed in these works
(e.g., Shaw et al. 2010; Trac et al. 2011; Sun et al. 2011; Shirokoff et al. 2011; Reichardt et al. 2011).

\section{Baryon budget in galaxy groups}

\subsection{X-ray gas fraction}

The halo gas fraction, or more precisely the likelihood function $p (f_{\rm gas} | M, r, z)$,
contains imprints of the thermodynamic history of the hot gas and is critical for using clusters to constrain
cosmology (e.g., Allen et al. 2002; V09).
The gas fraction usually increases with
radius (e.g., Sanderson et al. 2003; V06) and we focus on gas fraction within $r_{2500}$
(or $f_{\rm gas,~2500}$) and $r_{500}$ (or $f_{\rm gas,~500}$).
Results from the \rosat\ and \asca\ data showed that gas fraction generally increases
with mass (e.g., Sanderson et al. 2003) but uncertainty is large from the substantial
uncertainty on the total mass because of the little constrained temperature profiles
at large radii. \chandra\ and \xmm\ data provide more robust constraints
on gas fractions (e.g., V06, G07 and S09).
Groups have low gas fractions within $r_{2500}$, with a mass dependence of $M_{\rm HSE,~500}^{\sim 0.3}$ (S09).
The $f_{\rm gas,~500}$ difference between groups and clusters is smaller, with a mass dependence
of $M_{\rm HSE,~500}^{\sim 0.14 - 0.2}$ (S09, P09, also see Fig.~3).
Simulations with AGN heating generally fit the data better
(e.g., Puchwein et al. 2008; Bower et al. 2008; McCarthy et al. 2010; Fabjan et al. 2010),
which shows the vulnerability of the group gas to strong heating events.
It is essential to extend the mass coverage of the halo gas fraction relation and
better constrain the scatter. One recent interesting result is from Humphrey et al. (2011)
who studied an isolated galaxy (NGC~720) that is about ten times less massive than the
lowest-mass group in the S09 sample. As shown in Fig.~11 of Humphrey et al. (2011),
NGC~720's gas fraction follows the extrapolation of the cluster/group gas fraction relation
shown in Fig.~3 (the black solid line).

The gas fraction at $r > r_{500}$ depends on the gas density profile there.
For $M_{500}$ = 10$^{13}$ h$^{-1}$ M$_{\odot}$, $f_{\rm gas,~500} \sim 0.060$
from Fig.~3. Assuming $n_{\rm e} = 9\times10^{-5}$ cm$^{-3}$ (Fig.~2)
and a constant density slope ($n \propto r^{- 3 \beta}$) between $r_{500}$ and 2 $r_{500}$,
$f_{\rm gas}$ (@ 2 $r_{500}$) = 0.127 for $\beta$ = 0.4, 0.103 for $\beta$ = 2/3 and 0.093 for $\beta$ = 0.8.
Clusters have an average $\beta$ of 0.7 - 0.8 at $r_{500}$ (e.g.,
Vikhlinin et al. 1999; V06; Croston et al. 2008a; Eckert et al. 2011b)
and the average $\beta$ further increases to $\sim$ 0.9 at $\sim r_{200}$
(e.g., Ettori et al. 2009; Eckert et al. 2011b).
For groups, S09 measured an average $\beta$ of $\sim$ 0.6 at $r_{500}$
for $kT_{500} <$ 2 keV systems but there is a hint of correlation between
$\beta$ and $kT_{500}$. Rasmussen et al. (2010) measured $\beta = 0.45\pm0.05$ at $r_{500}$
for a 0.9 keV group. The density slope at $r > r_{500}$ is little known for groups, while
deeper X-ray observations and stacking of the existing \chandra\ and \xmm\ data should
help. Such kind of analysis will also constrain the clumpiness of the hot gas at $r > r_{500}$. 
The X-ray emission probes the square of the gas density so the small-scale density
inhomogeneity would bias the derived gas mass high.
One recent example of deep X-ray observations is with the
\suzaku\ data that Humphrey et al. (2012) traced the gas emission in a fossil group
to $\sim$ 2 $r_{500}$ and measured $\beta = 0.59\pm0.09$ between $r_{500}$ and 2 $r_{500}$,
and a gas fraction of 0.14$\pm$0.02 h$_{70}^{-1.5}$ within 2 $r_{500}$, which shows no needs
for gas clumping within 2 $r_{500}$.

Non-X-ray selected groups and clusters are on average X-ray fainter than X-ray
selected systems with similar values of mass proxies (e.g., optical richness,
integrated optical luminosity and velocity dispersion) (e.g., Rasmussen et al. 2006;
Popesso et al. 2007; Rykoff et al. 2008a; Dietrich et al. 2009).
The X-ray luminosity - mass relation also has a large scatter
(e.g., Rykoff et al. 2008b).
The effect of sample selection on the halo gas fraction is still unclear,
as the bulk of the gas mass is from X-ray faint regions at $r > r_{2500}$.
A recent work by Dai et al. (2010) derived very low gas fraction for $kT <$ 1.2 keV groups,
from stacking the \rosat\ all-sky survey data of the 2MASS selected clusters and groups.
The same analysis also derived a similar gas fraction for $kT \sim$ 1.7 keV systems as the S09
result. As the stacked structure functions of the gas (e.g., $\beta$ or the density slope
at large radii) are similar for $kT <$ 2.5 keV systems
from Dai et al. (2010), low gas fraction would come from low density normalization. 
It is known that optically (or NIR) selected systems are contaminated by projection effect
(especially for $M_{500} < 10^{14}$ M$_{\odot}$ systems, e.g., Koester et al. 2007).
Young systems collapsing for the first time (e.g., Rasmussen et al. 2006) is another
concern.
Stacking is also a tricky procedure that needs to be better calibrated.
On the other hand, the thermal SZ power spectrum provides another constraint
to the halo gas fraction, as its amplitude scales roughly as the square of halo gas fraction
(also see \S4). As discussed in Shaw et al. (2010), Trac et al. (2011) and Sun et al. (2011),
the current difference between the observed
SZ power spectrum and the models can be largely reduced by non-thermal pressure
support in groups, without the need to significantly reduce the halo gas fraction
derived from the X-ray samples in the low-mass end.

Besides X-ray observations, indirect constraints on the gas (or pressure) content in
groups may come from the role of ram pressure stripping on the star formation history
of group galaxies (e.g., von der Linden et al. 2010; Wetzel et al. 2011) and the bent radio galaxies
in groups (e.g., Freeland \& Wilcots 2011).

\subsection{Stellar mass fraction}

While the stellar mass fraction in rich clusters is small, it may be comparable
to the gas mass fraction in groups, especially if the mass of the intracluster light
\footnote{We still use ``intracluster light'' to refer the unbound stars between group galaxies
in this paper.}
is accounted for (e.g., Gonzalez et al. 2007; Giodini et al. 2009; McGee \& Balogh 2010).
However, a recent analysis of the COSMOS groups
(Leauthaud et al. 2012) suggests a significantly smaller stellar mass fraction
for galaxy groups than previous results (also see Fig.~3), with the difference
attributed to the smaller M/L ratios for the satellite galaxies than the central
galaxy and the assumed initial mass function.
The samples of Lin et al. (2003) and Gonzalez et al. (2007) do not cover low-mass groups.
From Giodini et al. (2009) and Leauthaud et al. (2012), the stellar mass fraction
is $\sim$ 0.069 and $\sim$ 0.02 within $r_{500}$ for $M_{500}$ = 10$^{13}$ h$^{-1}$ M$_{\odot}$
respectively.
Therefore, the Leauthaud et al. (2012) results imply a group baryon fraction that is
significantly lower than the universal baryon fraction at $r_{500}$ and it may remain
so at $r_{\rm vir}$ (or even 2 $r_{500}$), unless the gas density profile is very shallow
at $r > r_{500}$ (e.g., $\beta \leq$ 0.4).
However, the intracluster light in groups, which is not included in the analysis of
Lin et al. (2003), Giodini et al. (2009) and Leauthaud et al. (2012), still needs to be better
constrained. One recent analysis with SNIa suggests similar stellar mass between
the intracluster light and the galaxies in groups (McGee \& Balogh 2010).

%%%%%%%%%%%%%%%%%%%%%%%%%%%%%%%%%%%%%%%%%%%%%%%%%
\begin{figure}[t]
\centerline{\includegraphics[width=0.4\textwidth,angle=270]{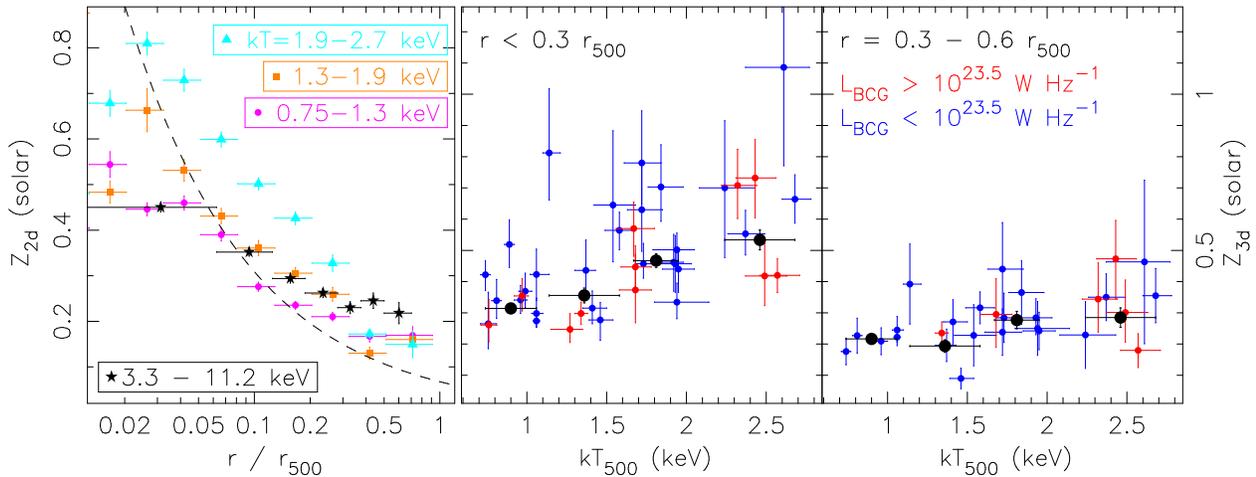}}
\caption{{\em Left}: The stacked projected abundance profiles for 39
groups from S09 in three temperature bins and for 48 clusters from
Leccardi \& Molendi (2008b).
The dashed line is the best-fit of the projected iron abundance profile from
Rasmussen \& Ponman (2009) for cool-core groups with a median temperature of 1.06 keV.
The hot gas in groups is iron poorer than the hot gas in clusters at $r$ = 0.3- 0.7 $r_{500}$.
{\em Middle \& Right}:
The gas-mass-weighted, deprojected abundance of 39 groups at $r < 0.3$ $r_{500}$
and $r = 0.3 - 0.6$ $r_{500}$ from S09. The red points are for 11 groups where the
brightest cluster galaxy (BCG) hosts a radio AGN with $L_{\rm 1.4 GHz} > 10^{23.5}$ W Hz$^{-1}$,
while the blue points are for 28 groups hosting a weaker radio AGN at the center.
Black points show the weighted averages in four bins. On average, lower-mass groups
have less metals at the center than higher-mass groups (also see Rasmussen \& Ponman 2009).
Larger samples than the S09 one are required to examine the connection between
the radio AGN activity and the abundance distribution.
}
\end{figure}
%%%%%%%%%%%%%%%%%%%%%%%%%%%%%%%%%%%%%%%%%%%%%%%%%

\section{Are the hot gas in galaxy groups metal poor?}

The global properties and distribution of heavy elements in the hot gas provide vital
constraints on the transferring mechanisms of metals from galaxies to the hot gas, and the
role of mergers, galactic winds and AGN outbursts (e.g., Renzini 1997; Buote 2000b; Baumgartner et al. 2005;
Rasmussen \& Ponman 2009; Fabjan et al. 2010; McCarthy et al. 2010).
Early X-ray observations constrained the total, emission-weighted abundance for groups and
clusters (e.g., Renzini 1997; Baumgartner et al. 2005). The global results imply that the
hot gas in groups have a much lower iron mass-to-light ratio than the hot gas in clusters
(e.g., Renzini 1997). Moreover, the hot gas in groups appears to be poor on iron and silicon
(e.g., Baumgartner et al. 2005). However, metals do not distribute uniformly in the hot gas
and the bulk of the gas mass is from the X-ray faint outskirts. Thus, it is necessary to
re-examine these results with spatially resolved spectroscopy. The \chandra\ and \xmm\ data
allow constraints on the distribution of heavy elements (radial profiles and abundance map),
which are briefly summarized in this section.
As different works used different solar abundance tables, we have converted all results to the
basis of the solar abundance table from Anders \& Grevesse (1989).

Fig.~4 shows the derived abundance of the hot gas in groups from S09, in three
temperature bins. Abundance was derived from the spectral fits with the APEC model. Deprojected
abundance profile was derived for each group. Along with the gas mass profile, the
gas-mass-weighted abundance is derived for 39 groups.
The left panel of Fig.~4 also shows the stacked projected abundance profiles for 48 $z=0.1-0.3$
clusters from the \xmm\ data (Leccardi \& Molendi 2008b), as well as the best-fit of the projected iron
profile from Rasmussen \& Ponman (2009) for cool-core groups (dashed line).
The projected abundance values from S09 are somewhat higher than that from
Rasmussen \& Ponman (2009) at $r >$ 0.3 $r_{500}$. The S09 results are consistent with the results from
Johnson et al. (2009) (a sample of 28 groups with a median temperature of 1.04 keV from the \xmm\
data) and are also consistent with the \suzaku\ results for five $\sim$ 1 keV groups (a median iron
abundance of $\sim$ 0.2 at $r \sim$ 0.35 $r_{500}$, Murakami et al. 2011; note the different solar
abundance tables used in different works).
Buote et al. (2004) also reported a low iron abundance ($\sim$ 0.1 solar) in a southern portion
of the (0.3 - 0.6) $r_{500}$ sector for NGC~5044, but the iron abundance in another direction of
the (0.3 - 0.45) $r_{500}$ sector from the \suzaku\ data is twice higher (Komiyama et al. 2009).
Humphrey et al. (2012) measured an emission-weighted iron abundance
of 0.19$\pm$0.05 solar at 0.3 $r_{500}$ - 1.5 $r_{500}$ for a fossil group.
We conclude that the iron abundance of the group gas is $\sim$ 30\% lower than
that of the cluster gas at $r$ = (0.3 - 0.7) $r_{500}$. 
Beyond the radius of 0.7 $r_{500}$, few groups have abundance constrained well.

For $kT_{500}$ = 0.8 - 2.7 keV systems, the average iron abundance within $r \sim 0.3$~$r_{500}$
increases with mass, but the trend almost disappears at $r$ = (0.3 - 0.7) $r_{500}$
(Rasmussen \& Ponman 2009 and Fig.~4). As shown in the left panel of Fig.~4, the hot gas in the
hottest groups seem to be the most iron rich within $r \sim 0.3$~$r_{500}$, even compared
with the hot gas in clusters.
This average abundance trend with the system temperature is similar to what was found from
the integrated, emission-weighted abundance
(Baumgartner et al. 2005), although the change on the gas-mass-weighted abundance with temperature
is smaller than the change on the emission-weighted abundance. While part of the abundance
excess for hot groups may be explained by the ``inverse iron bias'' (Rasia et al. 2008),
cool groups appear to be the most iron deficient within $r \sim 0.3$~$r_{500}$.
The metal deficiency within 0.3 $r_{500}$ is mainly contributed by regions beyond the group
cool cores (with an average radius of $\sim$ 0.1 $r_{500}$, see \S7.1) so it should not
be affected much by the ``iron bias'' (Buote 2000a).
The metal deficiency is also present for products from both SN Ia and SN II (Rasmussen \& Ponman 2009).
Therefore, within $r \sim 0.3$~$r_{500}$ groups with lower mass have weaker ability to retain
metal than the hotter groups, although the stellar mass fraction increases with decreasing system mass.
Possible scenarios to explain the observational trend include metal loss from AGN outbursts
in early stages or late stages, galactic winds and metal release efficiency
(e.g., Rasmussen \& Ponman 2009; Fabjan et al. 2010). 
We also attempted to look for the correlation between the abundance scatter and the
scatter on gas fraction and entropy at $r_{2500}$, but failed to establish such
a connection from the S09 data.

The group abundance profiles are generally centrally peaked with a near
solar abundance at the center (e.g., Buote 2000a; Buote 2000b;
Rasmussen \& Ponman 2009; Johnson et al. 2009; Fig. 4).
Johnson et al. (2011) divided their groups into cool cores and non cool cores, based on the central temperature drop
(see \S7.1 for the discussions on the group cool core). Groups with non cool
cores have flat abundance profiles. 
The results on the radial distribution of the $\alpha$/Fe ratios are not converged yet.
Rasmussen \& Ponman (2009) derived a rising Si / Fe profile at $r >$ 0.3 $r_{500}$,
while the \suzaku\ results are consistent with a flat $\alpha$/Fe ratios
profile to $r \sim$ 0.5 $r_{500}$ for Si, O, S and Mg (e.g., Sato et al. 2010; Murakami et al. 2011).

Looking into the future, it is important to have abundance well constrained (for both
iron and the $\alpha$/Fe ratios) for a sizable sample of groups to $r >$ 2/3 $r_{500}$
and extend the relations to groups with lower mass than what is discussed here.
Does the trend of lower abundance
for lower mass halos continue? What is the role of the late-stage AGN in transporting
metals in groups? How significantly is the hot gas clumped at large radii of groups
(e.g., $r > r_{500}$) and does the clumping bias the measured abundance low like
the ``iron bias'' in the core? Where are the missing metals in groups?
Evolution of the abundance content in groups can also be explored with the future data.

\section{Cool cores and AGN heating in galaxy groups}

\subsection{X-ray cool cores in groups}

The definition of cool cores is by no means trivial. Different papers may use
different definitions (e.g., Sanderson et al. 2006; O'Hara et al. 2006;
Chen et al. 2007; P09; Hudson et al. 2010), which can cause confusion.
For nearby systems where inner temperature profiles can be constrained, there
are three kinds of widely used definitions.
The most common definition uses the central cooling time. However, one has to define
a cooling time threshold and an inner radius where cooling time is measured.
There is no consensus so far on the definition. Cooling time thresholds of 1 - 11 Gyr
(for H$_{0}$ = 73 km s$^{-1}$ Mpc$^{-1}$) have been used to define cool cores
(e.g., O'Hara et al. 2006; Chen et al. 2007; P09; Hudson et al. 2010),
which inevitably affects
the derived cool core fractions (also affected by the sample bias, see Eckert et al. 2011a).
Recognizing the existence of clusters in the intermediate state, Hudson et al. (2010) defined three
classes of cores: strong cool cores (cooling time $<$ 1 Gyr), weak cool cores
(1 Gyr $<$ cooling time $<$ 7.7 Gyr) and non cool cores (cooling time $>$ 7.7 Gyr). 
The manifestations of cooling and star formation (optical emission-line filaments
and blue stellar cores) only happen in strong cool cores (e.g., Rafferty et al. 2008; Cavagnolo et al. 2008).
The choice of the inner radius where cooling time is measured is also tricky. 
Non cool cores generally have flat density and cooling time distribution at the center,
while many strong cool cores and some weak cool cores have continuously decreasing cooling
time profiles toward the center. In the literature, the inner radius is usually defined
relative to the virial radius (or $r_{500}$) but the cooling time profiles are not
similar with the scaled radius.
Hudson et al. (2010) measured the central cooling time at 0.004 $r_{500}$.
For the REXCESS clusters, the central density is derived from a $\beta$-model fit to
the deconvolved, deprojected density profile interior to 0.03 $r_{500}$.
In Fig.~5, we plot the radii at cooling time of 1 Gyr, 4 Gyr and 10 Gyr vs. the system temperature
for groups and clusters from S09, P09, V09, Maughan et al. (2008) and Rasmussen et al. (2010).
The average scaling relations of the cooling radii are flatter than the relation for the
scaled radius. If groups and clusters are self-similar, the gas density only depends on
the scaled radius. In cool cores, typically $n_{\rm e} \propto r^{-1}$
(e.g., Croston et al. 2008a; S09). Assuming an abundance of 0.6 solar,
the cooling radius - temperature relation is shown as the dashed-dotted line in Fig.~5,
which predicts group cool cores that are too large. As it is known that group cool cores
are also poor in hot gas compared with the cluster cool cores (\S4), an adjustment with
$n_{\rm e} \propto T^{0.6}$ (Table 1) better fits the data (the dashed line in Fig.~5).

%%%%%%%%%%%%%%%%%%%%%%%%%%%%%%%%%%%%%%%%%%%%%%%%%
\begin{figure}[t]
\centerline{\includegraphics[width=0.48\textwidth,angle=270]{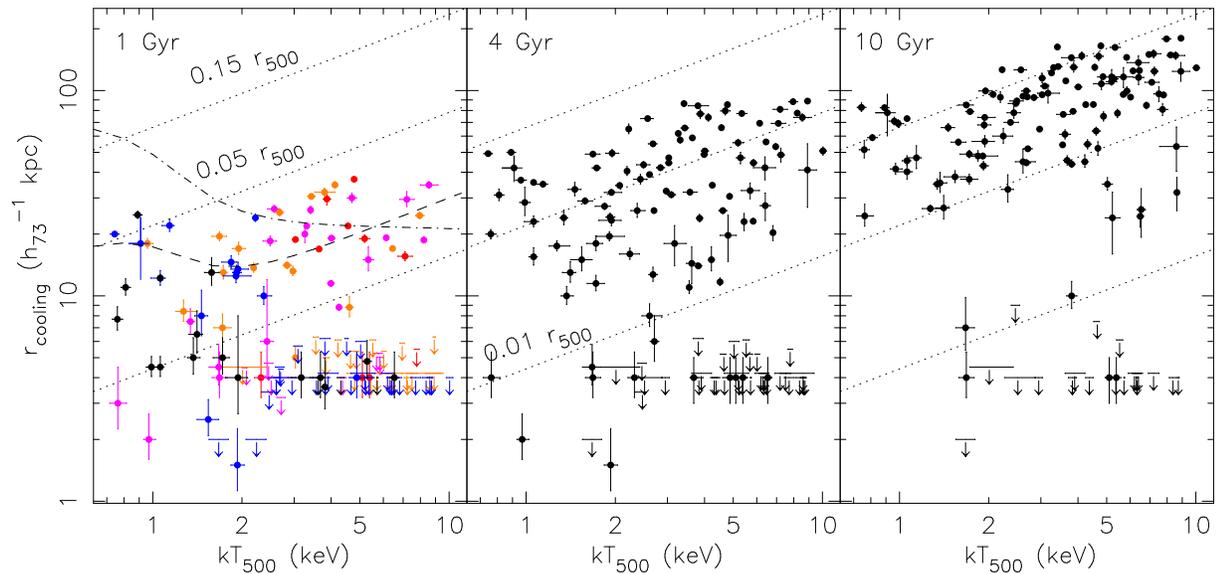}}
\caption{The radii at cooling time of 1 Gyr, 4 Gyr and 10 Gyr for 139 groups
and clusters from the S09 group sample, the REXCESS sample
(Croston et al. 2008a; P09), the \chandra\ cluster samples (Maughan et al. 2008; V09)
and Rasmussen et al. (2010) (one group).
The cooling radii of the \chandra\ clusters are from {\em ACCEPT} (Cavagnolo et al. 2009).
The cooling time definition is the same as Hudson et al. (2010).
The points in the 1 Gyr plot are colorized with the 1.4 GHz luminosity of the BCG:
red for $> 10^{25}$ W Hz$^{-1}$, magenta for $10^{24} - 10^{25}$ W Hz$^{-1}$,
orange for $10^{23} - 10^{24}$ W Hz$^{-1}$, blue for $10^{22} - 10^{23}$ W Hz$^{-1}$ and
black for $< 10^{22}$ W Hz$^{-1}$. Clusters with strong cool cores are always radio active
($L_{\rm 1.4 GHz} > 10^{23}$ W Hz$^{-1}$), which is not true for groups.
For systems where the cooling time threshold is not observed, upper limits
are given, typically at 4 kpc for possible faint coronae (Sun et al. 2007; Sun 2009).
Small coronae are included for the \chandra\ samples while the \xmm\ data for
the REXCESS sample do not have adequate spatial resolution.
While the 1 Gyr plot implies a bimodal distribution for cluster cool cores (large
cool cores + coronae), the distribution becomes continuous for groups.
The average cooling radius - $kT_{500}$ relations are flatter than the
relations for the scaled radius (dotted lines).
In the 1 Gyr plot, we show two models, the dashed-dotted line for the self-similar systems
(or gas density only depends on the scaled radius, not the system mass), and the dashed line for systems
with $n \propto T^{0.6}$ adjustment (see \S7.1 and \S4 for detail). In both cases, normalization
is simply adjusted to match the large cool cores in hot clusters.
The self-similar model predicts larger cool cores (defined by the cooling time) in groups
than cool cores in clusters, which is not observed.
This comparison again shows that the group cool cores are poorer in hot gas than
the cluster cool cores.
}
\end{figure}
%%%%%%%%%%%%%%%%%%%%%%%%%%%%%%%%%%%%%%%%%%%%%%%%%

%%%%%%%%%%%%%%%%%%%%%%%%%%%%%%%%%%%%%%%%%%%%%%%%%
\begin{figure}[t]
\centerline{\includegraphics[width=0.75\textwidth,angle=270]{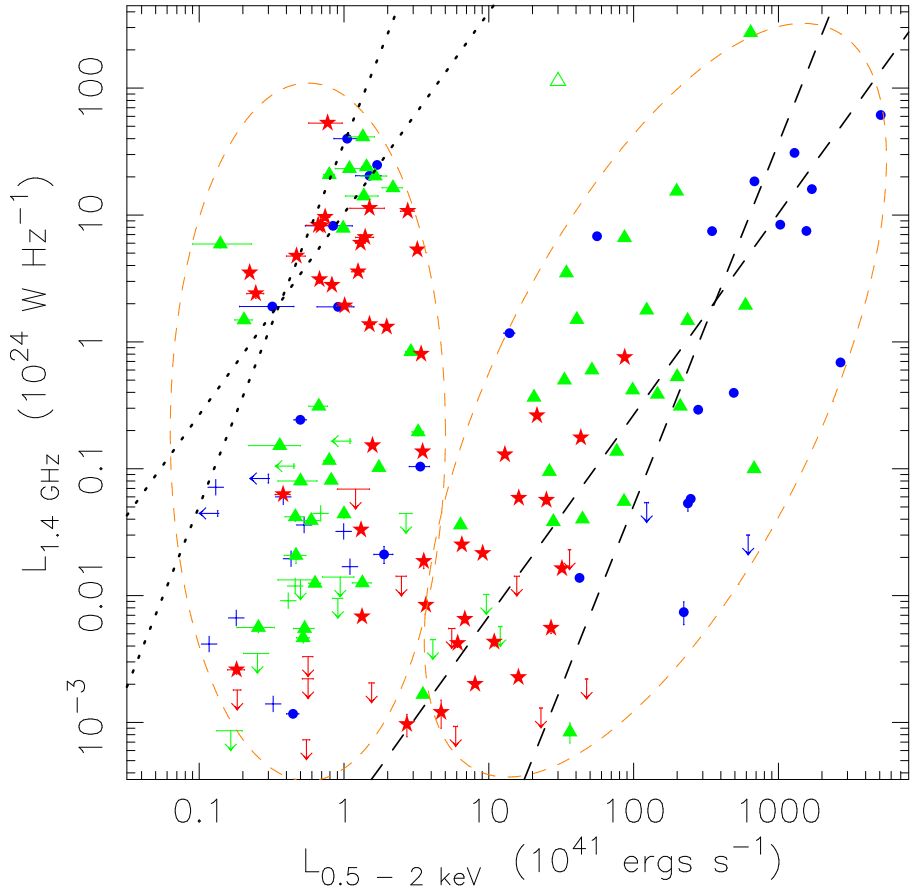}}
\caption{The rest-frame 0.5-2 keV luminosity of the cool core
(within a radius where the cooling time is 4 Gyr) of the BCG vs. the 1.4 GHz
luminosity of the BCG (an updated version of Fig.~1 of Sun 2009).
Red stars are $kT<$ 2 keV groups. Green triangles are $kT$ = 2 - 4 keV poor clusters.
Blue circles are $kT>$ 4 keV clusters.
Crosses represent upper limits in both axes. Two classes (large cool cores
and coronae) identified by Sun (2009) are enclosed by orange ellipses.
The two dashed lines on the right are the best-fit $L_{\rm 1.4 GHz} - P_{\rm cav}$
relations from Birzan et al. (2008) (the steeper one) and Sullivan et al. (2011a),
with an average bolometric correction factor of 2.5 used.
The heating lines roughly match the axis of the large cool core class
(see Sullivan et al. 2011a for discussions on the uncertainty of the cavity power relation),
which suggests that the large cool cores are AGN feedback regulated.
The two dotted lines on the left represent 0.1\% of the cavity heating power.
Coronae and small cool cores pose strong constraints to AGN heating
and feedback models.
Groups with radio AGN of $L_{\rm 1.4 GHz} > 10^{24}$ W Hz$^{-1}$ do not host
large, luminous cool cores, probably because these cores will be
over-heated and decoupled from the AGN feedback regulated cycles (see discussions
in Sun 2009).
}
\end{figure}
%%%%%%%%%%%%%%%%%%%%%%%%%%%%%%%%%%%%%%%%%%%%%%%%%

%%%%%%%%%%%%%%%%%%%%%%%%%%%%%%%%%%%%%%%%%%%%%%%%%
%\vspace{0.8cm}
\begin{figure}[t]
\centerline{\includegraphics[width=0.65\textwidth,angle=270]{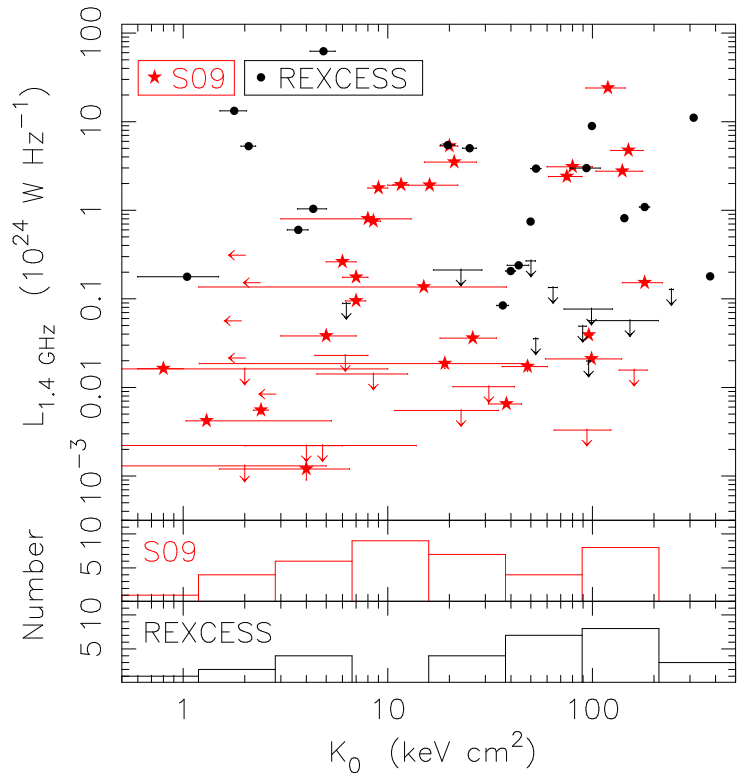}}
\caption{The central entropy ($K_{0}$) of the S09 groups in red and the
REXCESS clusters (Pratt et al. 2010) in black vs. the 1.4 GHz
luminosity of the BCG. Histograms for both samples are shown at the bottom panels.
A bigger sample than the S09 sample is required to examine the core entropy
distribution in groups.
}
\end{figure}
%%%%%%%%%%%%%%%%%%%%%%%%%%%%%%%%%%%%%%%%%%%%%%%%%

The choice of the inner radius is also complicated by the population of kpc-size cool cores (or coronae)
associated with the BCG in many non cool cores and weak cool cores (Sun et al. 2007; Sun 2009).
\footnote{Following Sun (2009), we call these small cool cores as coronae, typically
with a radius of several kpc. The classical cool cores are called large cool cores.}
Despite their small sizes, they carry enough fuel to power FR-I radio sources and indeed
many of them are associated with strong radio AGN (Sun 2009).
Coronae are not included in the classification of cluster cool cores
(Sanderson et al. 2006; O'Hara et al. 2006; P09; Hudson et al. 2010) and coronae are much
smaller than the large cool cores in clusters (Fig.~5). However, as the sizes and X-ray
luminosities of coronae correlate little with the system mass (Sun et al. 2007; Sun 2009),
they become inevitably prominent in low-mass systems.
Fig.~5 includes coronae and upper limits (in cases where coronae are not detected).
Evidently, the gap between large cool cores and coronae disappears for groups in the 1 Gyr plot.
This trend can also be seen in the X-ray luminosity of cool cores (Fig.~6) (updated from Sun 2009).
Therefore, the distribution of strong cool cores (in term of size and X-ray luminosity) may
become continuous in groups, compared with the likely bimodal distribution for clusters
(large cool cores vs. coronae).

The second definition of cool cores relies on the temperature drop at the core. 
For example, Sanderson et al. (2006) used the $kT (< 0.1 r_{500}) / kT (0.1-0.2 r_{500})$ ratio and
Johnson et al. (2009) used the $kT (< 0.05 r_{500}) / kT (0.1-0.3 r_{500})$ ratio.
Cool cores are defined if the ratio is less than unity.
For clusters, this kind of definition mainly selects strong cool cores as they generally
have temperature drops towards the center, while many weak cool cores defined by cooling time
lack significant temperature drops. For example, seven of eleven non cool cores defined
by Sanderson et al. (2006) are weak cool cores defined by Hudson et al. (2010).
For groups, many non cool cores defined by Johnson et al. (2009) have gas with short cooling time
around the center and the BCG coronae.
Under this definition, group cool cores are smaller than the cluster cool cores in term of
the scaled radius, as the group temperature profiles generally peak at smaller scaled
radius than the cluster profiles (e.g., Sun et al. 2003; V06; Rasmussen \& Ponman 2007; S09; also see \S4).
This definition may have problem for $kT <$ 0.6 keV systems, where the inner
temperature gradient is generally negative (e.g., Fukazawa et al. 2006;
Humphrey et al. 2006; Diehl \& Statler 2008).
One of the earliest examples is NGC~6482 (Khosroshahi et al. 2004).
Johnson et al. (2009) also suggested a lack of cool-core systems at $kT <$ 0.7 keV.
It remains to be seen whether there are $kT <$ 0.6 keV groups with a positive temperature gradient
at the center and whether strong radio AGN exist in such low-mass groups (Diehl \& Statler 2008).

The third definition uses the central entropy, if its distribution is bimodal
(see Cavagnolo et al. 2009, P09, Hudson et al. 2010 for discussions).
In Fig.~7, the central entropy (see definition in Cavagnolo et al. 2009) is
plotted with the 1.4 GHz luminosity of the BCG for the S09 sample
and the REXCESS sample (P09). The results for the group sample are
inconclusive, also because the S09 sample is biased towards the relaxed groups.
On the other hand, we emphasize that, even if the entropy bi-modality exists,
it means bimodal entropy distribution at scales beyond 10 kpc, not necessarily
a bimodal entropy distribution around the central black hole.
As shown by Sun (2009) on the ubiquity of the BCG coronae,
high entropy at 10 kpc or larger scales does not mean low radio activity of the nucleus,
which is also evident from Fig.~7. 
\footnote{The search criteria for radio sources in Cavagnolo et al. (2008)
missed large, amorphous sources for low-$z$ systems (see a note on the {\em ACCEPT} website)
so there are strong radio AGN in high $K_{0}$ clusters (Fig.~2 of Cavagnolo et al. 2008).
However, all of them have coronae at the center which are not included in the
$K_{0}$ determination (see Sun 2009).}

To sum up, there exist different definitions of cool cores in the literature so
comparison should be done with caution. If cool cores are defined by cooling time,
the large cool cores in groups are on average less dense (at the same scaled
radius) but larger (in terms of the scaled radius) than the large cool cores
in clusters (Fig.~5). On the other hand, while the cool core distribution (size and
X-ray luminosity) may be bimodal in clusters (large cool cores vs. coronae), the
distribution is more continuous in groups (Fig.~5 and 6). It is interesting to
explore whether mergers of group cool cores can result in cluster cool cores
(e.g., Motl et al. 2004), especially if coronae can be seeds of group cool cores.
It is unclear whether there is a bi-modality of the core entropy in groups, as larger and better
samples than the S09 one are required. It is also intriguing to study the gas cores
of low-mass groups ($kT_{500} <$ 0.6 keV), on e.g., the temperature profile and
the connection with the radio activity of the BCG.

\subsection{AGN heating in groups}

AGN heating is considered to be a crucial ingredient in structure formation and the
leading solution to the cooling flow problem.
The BCG stellar light only weakly depends on the halo mass
($L_{\rm Ks} \propto M^{\sim 0.26}$, Lin \& Mohr 2004) and the relation has large scatter.
Thus, the average strength of the radio AGN outbursts from BCG may only weakly
depend on the halo mass.
As the cool core luminosity decreases by a factor of $\sim$ 100 from clusters
to groups (e.g., Fig.~6), heating should dominate in groups and can change the
properties of the hot gas significantly.
By no means a complete review of AGN heating in groups (see e.g., Mathews \& Brighenti 2003;
McNamara \& Nulsen 2007; Gitti et al. 2012; McNamara et al. 2012),
this section only summarizes several difference of AGN heating in groups from in clusters. 

Few, if any, groups with large, strong cool cores host strong radio AGN.
As shown in Fig.~5, all 28 clusters ($kT_{500} > 2.5$ keV) with large, strong cool cores
(cooling time of $<$ 1 Gyr at 10 kpc) are radio active ($L_{\rm 1.4 GHz} > 10^{23}$ W Hz$^{-1}$),
which is not true for groups. Combined with Fig.~6, it appears
that groups with large, strong cool cores do not host strong radio AGN.
Those strong radio AGN in groups are only associated with small cool cores like coronae.
The group BCGs with large or small cool cores have similar stellar light (Sun 2009)
so the difference on the black hole mass should not be the main reason.
One possibility suggested by Sun (2009)
is the destroying of large group cool cores when strong radio outbursts are triggered.
While large cluster cool cores can ``contain'' and survive strong radio outbursts,
the bulk of group cool cores can be over-heated and may not survive strong radio outbursts.
For groups with large (e.g., 1 Gyr cooling radius of $>$ 15 kpc) cool cores,
the radio AGN heating needs to be gentle (e.g., Gaspari et al. 2011).

The strong radio AGN in groups still co-exist with small cool cores (or coronae,
see Fig.~6). The significance of coronae is to test the AGN feedback models in extreme
conditions. An outstanding question with AGN feedback is to connect physics spanning over
nine orders of magnitude in spatial scales, from the event horizon to Mpc scales.
Coronae or small cool cores appear isolated from the embedded hot gas with higher entropy
so the connection is likely broken over the boundary. Radio lobes and the bulk of the radio
emission are generally observed outside of coronae. Almost all heating energy should
indeed be released outside of coronae, for their ubiquitous survival (Sun 2009; Fig.~6).
It would require a fine tuning for a strong radio outburst to offset cooling inside a
small corona without completely destroying it (see Sun 2009 for more discussions). 
Therefore, Sun (2009) suggested that coronae are decoupled from the AGN feedback
regulated states (also see Sullivan et al. 2011b).

Groups have low gas fractions within $r_{2500}$ and the radio AGN outbursts have significant
impact on group properties (e.g., Croston et al. 2005; Sun 2009).
Therefore, groups that have experienced strong heating episodes would be X-ray
faint and are difficult to be studied in detail in X-rays. These systems
would also be underrepresented in group samples selected by X-ray cavities with the strong contrast.
Indeed, there are 22 groups ($kT < 2$ keV) with X-ray cavities studied in the samples by
Birzan et al. (2008), Cavagnolo et al. (2010), Dunn et al. (2010) and Sullivan et al. (2011a).
The median 1.4 GHz luminosity is only 10$^{22}$ W Hz$^{-1}$ and the most luminous radio AGN
is 1.3$\times10^{24}$ W Hz$^{-1}$ (see Fig.~6 for the range of $L_{\rm 1.4 GHz}$ for BCG).
Even in these samples without strong radio AGN, deep X-ray observations are sometime
required to constrain the full AGN mechanical power. One example is NGC~4261.
The AGN mechanical power derived from the deep \xmm\ data (Sullivan et al. 2011b) is over an order of
magnitude higher than that derived from the \chandra\ data (Cavagnolo et al. 2010). 
Perhaps the same reason also explains the lack of strong shocks detected in groups.
Many weak shocks have been detected in clusters (see McNamara \& Nulsen 2007; Gitti et al. 2012).
Most of these radio outbursts will generate strong shocks in groups (Mach number $>$ 2). However, all
reported shocks in groups are weak ones with Mach number of 1.5 - 1.7 (Gitti et al. 2010; Randall et al. 2011).
Strong outbursts may have blown out the group cool cores, if only a small central core is left.
Indeed the only two groups with weak shocks detected (HCG~62 and NGC~5813) have weak
radio AGN ($L_{\rm 1.4 GHz} \sim 1.9\times10^{21}$ W Hz$^{-1}$).
Deep X-ray observations will be required to reveal strong shocks in groups, especially
around strong radio AGN.

%%%%%%%%%%%%%%%%%%%%%%%%%%%%%%%%%%%%%%%%%%%%%%%%%
\begin{figure}[t]
\centerline{\includegraphics[width=1.02\textwidth]{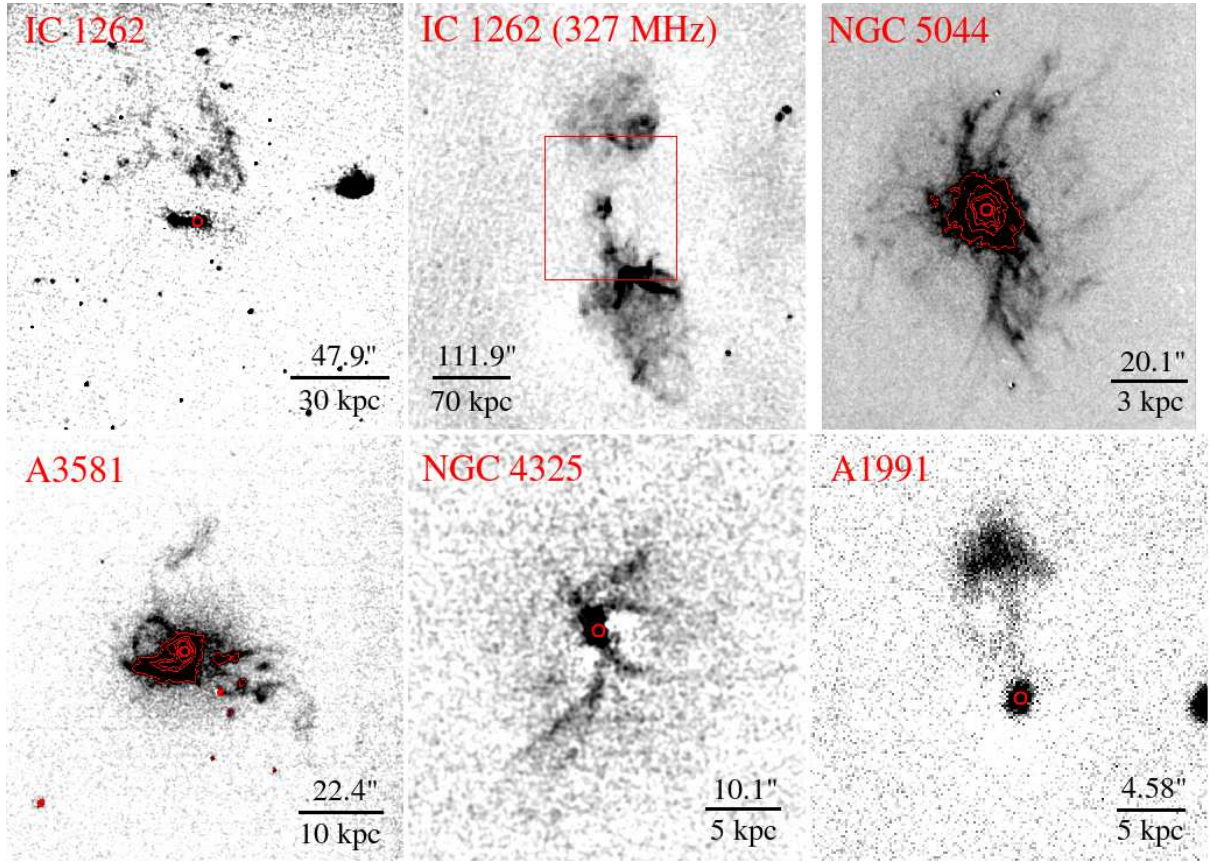}}
\caption{The net H$\alpha$+[NII] images of cool cores in IC~1262, NGC~5044,
A3581, NGC~4325 and A1991, from the {\em SOAR} 4.1m or the {\em APO} 3.5m telescope.
The inner structure
of the optical nebulae in NGC~5044 and A3581 is also shown with the red contours.
The \vla\ 327 MHz image of IC~1262 is also shown (courtesy of F. Owen, J. Eilek
and W. Forman) and the red box is the optical region shown on the left.
The small red circle shows the location of the nucleus.
This montage shows the complexity of the optical nebulae in group cool
cores (also see McDonald et al. 2011), in scales of kpc - 50 kpc.
}
\end{figure}
%%%%%%%%%%%%%%%%%%%%%%%%%%%%%%%%%%%%%%%%%%%%%%%%%

To sum up, radio AGN outbursts have larger impact on group cool cores than on cluster
cool cores. Groups with large, strong cool cores do not host strong radio AGN (Fig.~6),
probably because these extended cool cores can be easily over-heated during strong
heating events. The strong radio AGN in groups only co-exist with small cool cores like
coronae. Cooling in these small cool cores is still strong and it is likely they are
decoupled from the AGN feedback regulated state.
Therefore, group cool cores provide extreme conditions
to test the AGN feedback models (e.g., McCarthy et al. 2010; McCarthy et al. 2011; Gaspari et al. 2011).
The high X-ray surface brightness core sample for groups produces a biased view of the AGN
heating in groups. Deep X-ray observations of groups with strong radio AGN are required for
a more complete view.

\subsection{Cold gas in group cool cores}

The searches for cold gas in cool cores (e.g., optical emission-line nebulae, rotational
and vibrational H$_{2}$ emission, FIR atomic lines and CO emission) have been focused
on clusters, for their higher classic mass deposition rates. However, such kind of work is
also important in group cool cores. Heat conduction is much weaker in groups than in clusters.
Even without magnetic field, the mean free path of particles becomes small in group cool
cores (e.g., 11 pc for $kT$=0.6 keV and $n_{\rm e}$=10$^{-2}$ cm$^{-3}$)
that it is still unclear how it compares with the width of the optical filaments.
Groups with large, strong cool cores do not host strong radio AGN (\S7.2) so the cosmic
ray energy density may be smaller. Therefore, the group cool core is a critical environment
to study the optical emission-line filaments and the ionization mechanism (e.g., Ferland et al. 2009).
Indeed many large group cool cores host optical emission-line nebulae
(e.g., Crawford et al. 1999; McDonald et al. 2011).
Fig.~8 shows the H$\alpha$+[NII] images of five nearby groups from our SOAR/APO project
on group cool cores. Many interesting features can be seen, including the one-sided H$\alpha$+[NII]
filaments/complex in IC~1262 that may come from the lifted cold gas by the AGN outburst
and narrow radial filaments in NGC~5044, A3581 and NGC~4325.
The detection frequency of optical emission-line nebulae in group cool cores
is similar to that in cluster cool cores (e.g., McDonald et al. 2011) and
the optical filaments are generally associated with regions with enhanced
cooling in X-rays (e.g., McDonald et al. 2011; Randall et al. 2011).
Almost all the published works on the CO emission, FIR atomic lines, vibrational and
rotational H$_{2}$ emission is on cluster cool cores, but more work on group cool cores
will be important.

\section{Important knowns and unknowns}

The new generation data from \chandra, \xmm\ and \suzaku\ have already significantly
improved our understanding on the hot gas in galaxy groups.
The following is a summary on what has been discussed in this paper:
1) Galaxy groups are not scaled-down versions of rich clusters following self-similar
relations. Particularly, groups are poor in hot gas within $r_{2500}$ (including the cool
cores), which explains
their faint X-ray luminosities and high gas entropy. On the other hand, groups
not necessarily have low gas fraction at $r > r_{2500}$. Without mass-selected samples, we have to
combine sample studies from different selection functions with deep or stacked observations
on the group outskirts to better constrain the halo gas fraction and the scatter.
The recent advances on the measurement of the thermal SZ power spectrum open a new window into
groups. The stellar mass fraction in groups is still uncertain to at least a factor of two, especially
since the mass fraction of the intracluster light is poorly constrained in groups.
With all these uncertainties, the baryon fraction in groups is not well determined within
$r_{500}$ or $r_{\rm vir}$, although most recent results suggest a deficiency of baryons compared
with the universal baryon fraction within $r_{500}$.
2) The hot gas in galaxy groups are also iron poor at 0.4 - 0.7 $r_{500}$, compared with the
hot gas in clusters.
The iron abundance of the inner regions ($r <$ 0.3 $r_{500}$) also increases with mass
for groups, which is in conflict with the higher stellar mass fraction and the lower gas
mass fraction in lower mass groups. It is unclear whether the missing iron has been ejected to
large radii ($r >$ 0.7 $r_{500}$) or has not been efficiently released into the hot gas.
The radial distribution of the $\alpha$/Fe ratios still needs to be better constrained.
The role of the late-stage AGN in transporting metals in groups is unclear yet.
3) There are different definitions of cool cores in the literature so
comparison should be done with caution. If cool cores are defined by cooling time,
the large cool cores in groups are on average less dense (at the same scaled
radius) but larger (in terms of the scaled radius) than the large cool cores
in clusters. On the other hand, while the cool core distribution (size and
X-ray luminosity) may be bimodal in clusters (large cool cores vs. coronae), the
distribution is more continuous in groups.
Group cool cores are more vulnerable to radio AGN heating, which can be seen from the
lack of strong radio AGN in large group cool cores. 
Therefore, group cool cores provide ideal conditions
to test the AGN feedback models. The group cores hosting strong radio
AGN are X-ray faint so cavities and shocks become subtle features, but a more complete
picture of AGN heating in groups requires studies of these faint systems.

Many outstanding questions remain and a partial list includes:
1) How complete are X-ray surveys to uncover group halos?
The local halo mass function is well known. For X-ray selected
samples, once the X-ray luminosity function and the $L_{\rm X} - M$ relation are better
determined, the completeness of the X-ray survey to map the group halos can be constrained
through abundance matching.
2) What is the group gas fraction at $r > r_{500}$? Is the gas density - mass relation
still flat at $r > r_{500}$? How significant is gas
clumping at $r > r_{500}$ for groups? This can be explored with deeper observations
or stacking the existing \chandra, \xmm\ and \rosat\ data.
3) Are the single power-law scaling relations (e.g., for X-ray luminosity, entropy and gas fraction)
broken for groups with lower mass than what is discussed here?
The low-mass groups, e.g., $M_{500} < 10^{13} h^{-1}$ M$_{\odot}$,
are observationally challenging but important to study.
4) What is the stellar mass fraction and scatter in groups? What is the fraction of the stellar
mass in the intracluster light in groups? Are there missing baryons
within $r_{\rm vir}$ in groups?
5) Where is the missing iron in groups? Will we find the unaccounted metals at $r > r_{500}$?
What is the distribution of the $\alpha$/Fe ratios in groups? Do groups with lower
mass than discussed here have continuously lower iron abundance at $r <$ 0.3 $r_{500}$?
6) Are there any $kT <$ 0.6 keV groups with positive temperature gradient at the center?
7) Are there any large group cool cores with strong radio AGN at the center? Is the gap
in Fig.~6 occupied with any group cool cores? Presumably, such kind of
systems should exist at the early stage of radio AGN heating.
Can we find strong AGN shocks in groups?
8) Is there entropy bi-modality in group cores? Do non-cool-core groups without a BCG
corona exist?
9) Can decoupled group cool cores evolve back to large cool cores by merging?
10) Are there group merger shocks? Are there radio halos and relics in groups, since
the strength of the merger shocks should be little mass dependent?
11) The evolution of hot gas in galaxy groups, not discussed here, is clearly vital,
also for the SZ power spectrum.

\ack
We thank Judith Croston, Fabio Gastaldello, Fill Humphrey, Alexie Leauthaud,
Gabriel Pratt, Jesper Rasmussen and Eduardo Rozo for providing detail of their published results.
We are grateful to helpful comments and suggestions by Trevor Ponman,
Andrey Kravtsov, David Buote, Mark Voit and Alastair Sanderson.
The work has been supported in part by \chandra\ grants GO1-12159A, GO0-11145C and \xmm\ grant NNX09AQ01G.

\begin{appendix}

\section{The S09 and OP04 luminosity results}

S09 did not publish the results on the X-ray luminosities, which are included in this
paper. Here we present more detail on the luminosity results for the S09 groups.
Traditionally, only one conversion factor
from the observed count rate to flux is used to derive the global luminosity.
However, the conversion factor becomes sensitive to abundance at $kT < 2$ keV.
Given the observed temperature and abundance gradient in groups, we used the observed
profiles of temperature, abundance and surface brightness to do conversion in each
radial bin for the spectral analysis. There are 4 - 23 bins and the median is 9.
This approach was adopted also because the global, emission-weighted abundance
is not derived in our analysis. The incomplete coverage of the \chandra\ data at large radii prevents
a direct derivation of such a global value.
From our simulations, if only the global values of temperature and abundance are
used in the conversion, the average luminosities (bolometric or 0.5 - 2 keV) for
groups only change by $\sim$ 5\%, but the scatter would increase by 10\% - 15\%.
The shown luminosities of the S09 groups in Fig.~1 are derived from the \chandra\ data,
which usually requires extrapolation to the full area within $r_{500}$.
As most of the X-ray emission is from the central region, the required
extrapolation is always small. For the bolometric luminosity, the fraction of the total
flux covered by the \chandra\ data has a median of 83\%. In fact, 17 of the 43 groups
in S09 have \rosat\ PSPC data, including all seven groups with less than 70\%
flux coverage by the \chandra\ data. Assuming the same global properties, the \chandra\
luminosities and the PSPC luminosities agree well and the difference is always less than 4\%.
We can also compare the S09 luminosities to the values from OP04. There are six common
groups between S09 and OP04. After correcting the assumed spectral models and distance,
the average luminosity ratio between OP04 and S09 is 0.96$\pm$0.03.
There are four S09 groups in the 400 deg$^{2}$ survey (Burenin et al. 2007).
The S09 luminosities always agree with the PSPC values within 1$\sigma$.

Fig.~1 includes the \rosat\ sample from OP04 but the \rosat\ results
have larger uncertainty than the results from \chandra\ and \xmm.
Firstly, temperature can be biased high because of the unresolved point
sources, especially for faint groups with low surface brightness.
One example is NGC~1587, the second faintest 1 keV group in OP04.
The \chandra\ data instead revealed a temperature decrease by a factor of two
while the luminosity is little changed (Helsdon et al. 2005).
Secondly, PSPC data have little ability to constrain abundance, while OP04
used the best-fit abundance for the conversion factor. For 33 groups in
OP04, 7 have abundance of $>$ 0.9 solar and 7 have abundance
of $<$0.1 solar, which increases the scatter (e.g., the two outliers at $kT \sim 0.2$ keV
that OP04 used zero abundance).
Thirdly, the OP04 groups are typically at smaller $z$ than the S09
groups so the peculiar velocity will also increase the luminosity scatter.
While a full re-analysis of the OP04 sample with the \chandra\ or
the \xmm\ data is not available, we show the trend of changes for six OP04 groups
in Fig.~1. These changes are simply collected from literature (Helsdon et al. 2005;
Croston et al. 2008b; Johnson et al. 2009), or made with the surface brightness fluctuation
distance (Tonry et al. 2001), or assuming an average abundance of 0.25 solar
(for two outliers at $kT \sim 0.2$ keV). The changes are incomplete as the updated
system temperatures for four groups are unknown.

\section{Comparison of the gas fraction results}

In this appendix, the results of total mass and gas fraction from V06, V09, G07,
S09 and E11 are compared. There are five groups in V06 and one in V09. For simplicity, we use
V06 to represent this sample of six groups.
Since the common systems between any two of the V06, G07
and E11 samples are also in the S09 sample, the S09 sample is used as the reference.
All six groups in V06 are in S09. S09 analyzed all the \chandra\ data used in V06
and added a deep exposure for A262. There are 12 common systems between G07 and S09.
G07 analyzed both the \chandra\ and the \xmm\ data, when available. Gas properties
around $r_{2500}$ and beyond are mainly constrained by the \xmm\ data, when available.
There are 15 common groups between S09 and E11 and both works analyzed the same
\chandra\ data.

The comparison is mainly done at $r_{2500}$ as few groups have the X-ray HSE mass
derived to $r_{500}$ in V06, G07 and S09. The gas fraction at $r_{2500}$, $f_{\rm gas,~2500}$,
was first compared (Fig.~B1). The results from V06 and S09 on average
agree within 2\%. $f_{\rm gas,~2500}$ from G07 is on average $\sim$ 20\% higher than the
S09 results but the worst difference (A2717) is still within 3.5$\sigma$.
Most of the E11 values are within 20\% from the S09 results, but several groups have
very small $f_{\rm gas,~2500}$. For example, $f_{\rm gas,~2500}$ is 0.014$\pm$0.002,
0.047$\pm$0.003, 0.043$\pm$0.002 and 0.058$\pm$0.001 from E11, S09, V06 and G07
respectively, for the only common group in all samples, MKW4.
Fig.~B1 also shows the comparison on $r_{2500}$.
The results from V06, G07 and E11 are on average 4\%, 8\% and 9\%
lower than the S09 results respectively. For the E11 results, the
difference increases to 14\% if the four groups (MKW4, NGC~533, NGC~1550 and NGC~5129)
with low $f_{\rm gas,~2500}$ derived by E11 are excluded.

%%%%%%%%%%%%%%%%%%%%%%%%%%%%%%%%%%%%%%%%%%%%%%%%%
\begin{figure}
\vspace{-0.2cm}
\centerline{\includegraphics[width=0.81\textwidth,angle=270]{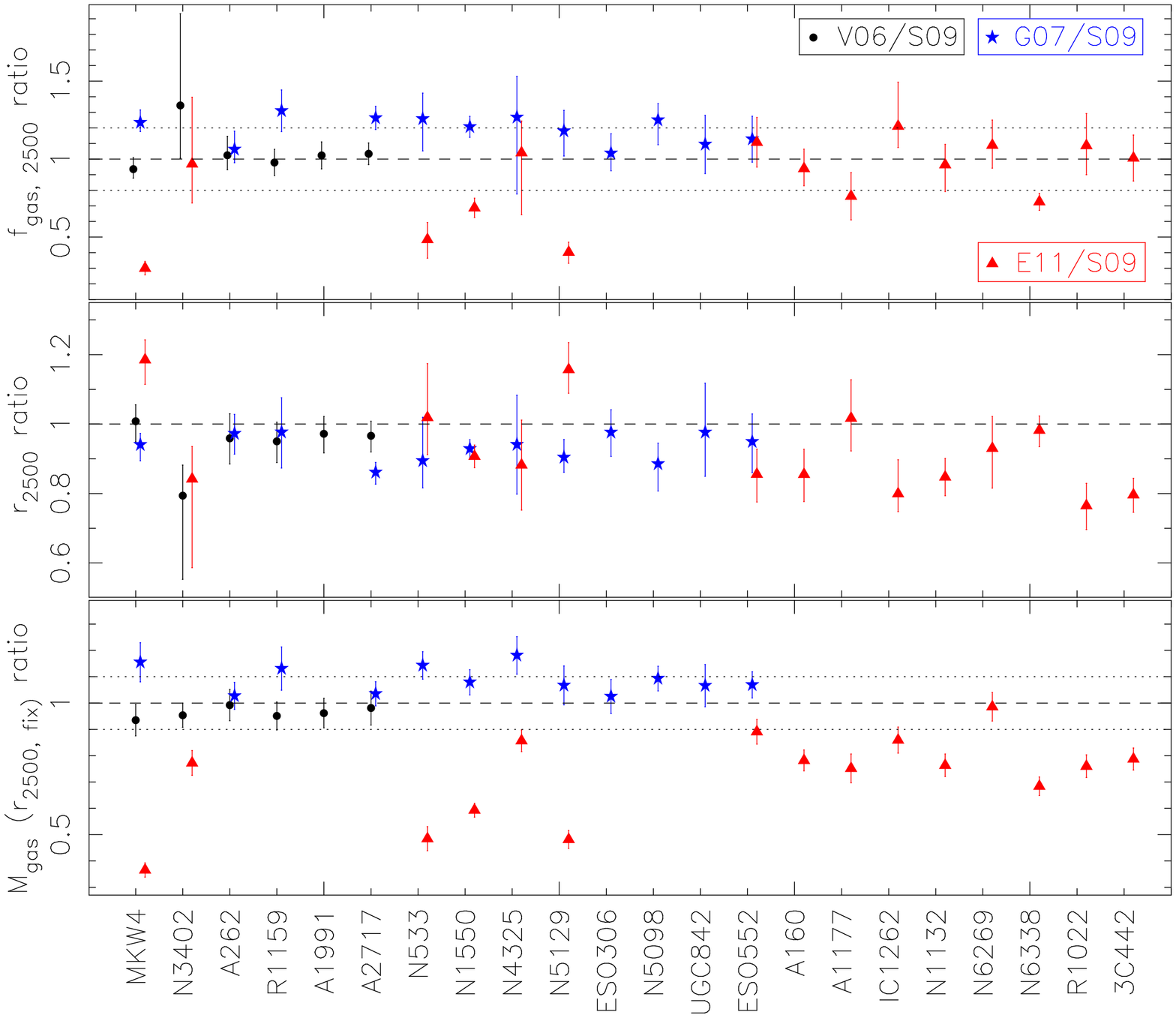}}
\vspace{-0.2cm}
\caption{{\em Top}: the ratios of $f_{\rm gas,~2500}$ by V06, G07 and E11 to
S09 for the same group. The dashed line is the line of equality and the two dotted
lines represent 20\% difference. While most results agree with the S09 results within 20\%,
E11 has several groups (MKW~4, NGC~533 and NGC~5129) with much lower $f_{\rm gas,~2500}$
than the values from the other work.
To further understand the difference, we compare $r_{2500}$ (or $M_{\rm HSE,~2500}$) and $M_{\rm gas}$
at fixed radius.
{\em Middle}: the ratios of $r_{2500}$ by V06, G07 and E11 to
S09 for the same group. The dashed line is the line of equality.
{\em Bottom}: the ratios of $M_{\rm gas}$ at the fixed $r_{2500}$ by V06, G07 and
E11 to S09 for the same group. The dashed line is the line of equality and the two dotted
lines represent 10\% difference. For the G07 results, the ratio is derived at the $r_{2500}$ derived
by G07, as we only have the S09 gas mass profile. The ratios are calculated similarly
for the V06 and the E11 results, at the $r_{2500}$ derived by the respective work.
The results from V06 and G07 are on average 4\% lower and 8\% higher than the S09 results
respectively. The E11 results are on average 34\% lower than the S09 results.
Even if the four groups with low $f_{\rm gas,~2500}$ (MKW4, NGC~533, NGC~1550 and NGC~5129)
are removed, the rest of the E11 results are on average 20\% lower than the S09 results.
Thus, the V06, G07 and S09 results agree to $\sim$ 20\% on the gas fraction and the total mass,
and to $\sim$ 10\% on the gas mass. The difference between the \chandra\ results (V06 and S09)
and the G07 results (mainly derived with the \xmm\ data at the large radii) may come from the
\chandra/\xmm\ cross-calibration and the different methods used to derive the HSE mass,
but the good agreement on $M_{\rm gas}$ is encouraging. On the other hand, E11 has some systems
with puzzling low gas mass. Even if these four groups with low gas mass are excluded, the E11 results have
$\sim$ 20\% lower gas mass at fixed radius and $\sim$ 30\% lower total mass than 
the results from the other two \chandra\ works.
}
\end{figure}
%%%%%%%%%%%%%%%%%%%%%%%%%%%%%%%%%%%%%%%%%%%%%%%%%

To better understand the difference, we compare the gas mass derived at a fixed
radius from different work. For the ratios between V06 (or G07, or E11) and S09,
this radius is the $r_{2500}$ derived by V06 (or G07, or E11).
We derived the S09 gas mass at this fixed radius for each comparison.
The results are also shown in Fig.~B1. The results from V06 and G07 are on average 4\% lower
and 8\% higher than the S09 results respectively. However, the E11 values are on average
20\% lower than the S09 results, even if the four groups with low $f_{\rm gas,~2500}$
are removed (these four have 52\% lower gas mass than the S09 results).
Therefore, the V06, G07 and S09 results agree to $\sim$ 20\% on the gas fraction and the total mass,
and to $\sim$ 10\% on the gas mass. The level of the difference between the \chandra\ results (V06 and S09)
and the G07 results (mainly the \xmm\ data at the large radii) is not surprising, considering
the current status of the cross-calibration between \chandra\ and \xmm\
(e.g., Nevalainen et al. 2010) and the different methods used to derive the HSE mass.
Nevertheless, the good agreement on the gas mass is encouraging.
On the other hand, the E11 results on the four groups with low gas mass, MKW4, NGC~533,
NGC~1550 and NGC~5129, are way off the results of other work. Even if these four groups
are excluded, the gas mass and the $M_{\rm HSE,~2500}$ by E11 are $\sim$ 20\% lower
and $\sim$ 30\% lower respectively than the results from the other \chandra\ work.

Similarly, E11 derived very small $f_{\rm gas,~500}$ (e.g., $< 0.03$) for some groups,
including MKW4, NGC~5129 and HCG97 with the best-fit values of $f_{\rm gas,~2500} > f_{\rm gas,~500}$,
in contradiction with the usual trend (gas fraction increasing with radius) and the
results from V06, G07 and S09 for the same groups. For example, the $f_{\rm gas,~500}$ ratio
between G07 and E11 for five common groups ranges from 0.9 to 10.2 with a median of 4.3
(0.107 vs. 0.025).
\footnote{The G07 results of $f_{\rm gas,~500}$ are from extrapolation, provided
by Fabio Gastaldello, while E11 also used extrapolation to derive $f_{\rm gas,~500}$.}
The $f_{\rm gas,~500}$ ratio between S09 and E11 for seven common groups ranges from
0.9 to 7.1 with a median of 1.5.
We also examined the difference on $r_{500}$ and the gas mass at fixed radius.
The conclusion is similar to what is drawn from the comparison at $r_{2500}$.

There are several concerns about the E11 work.
The {\em first} issue is the coverage of the group emission by the \chandra\ data. Derivations of
the group properties to $r_{500}$ (especially the X-ray HSE mass) require a good coverage
of the $r < r_{500}$ regions by the data, which is typically difficult for nearby groups
(e.g., $z <$ 0.02-0.03) with \chandra\ or even \xmm\ (e.g., $r_{500}$ = 15$'$ and 22$'$ for $kT_{500}$ = 1 keV
and 2 keV systems respectively at $z$=0.025).
Among the 43 groups in the S09 sample, only 11 groups (tier I groups in S09)
have gas properties derived to $r_{500}$ and another 12 groups (tier II groups in S09)
have gas properties derived to $r_{1000} < r < r_{500}$.
For better constraints on the gas emission at large radii, S09 also included
the \rosat\ PSPC data for 17 groups, including five tier I groups
and three tier II groups. S09 also presented the group properties to $r_{500}$
for tier II groups as only small radial extrapolation is required
($r_{1000} / r_{500} \sim$ 0.74 for the S09 groups).
As shown in $\S$2, the E11 groups are on average closer
than the S09 groups. The \chandra\ data only reach $\sim$ 0.6 $r_{500}$ on average and
the covered area beyond $r_{2500}$ is generally very small.
Yet E11 derived the group properties (including the total HSE mass)
to $r_{500}$ for all 26 groups.
There are 15 common groups between S09 and E11 and the same \chandra\ data were
analyzed. S09 only listed three groups in tier I and
four groups in tier II, even with the help of the PSPC data for 7 of 15 groups that
the E11 analysis did not use. 
Similarly, G07 only derived the group properties to $r_{500}$ for 3 out of 16 groups in their
sample and emphasized the bias with extrapolation.
There are 6 common groups between G07 and E11. All six groups have the \xmm\ data analyzed in G07
but none of them has the group properties derived to $r_{500}$ in G07.
For five groups, the data used in E11 have poorer coverage for $r > r_{2500}$
regions than the data used in G07.
Thus, the E11 results at $r > r_{2500}$ heavily rely on extrapolation.
The {\em second} issue is the background analysis.
Both the G07 and the S09 work have extensive discussions on the X-ray background,
as it is the key to have robust constraints on the gas properties beyond the core.
S09 also listed the derived fluxes of the local X-ray background. The derived local
X-ray background also compares well with the expectations from unresolved point sources
and the soft X-ray background observed by \rosat.
E11 did not present any details on their background analysis and the
background adjustment from the blank sky background they used.
The {\em third} issue is the modeling to derive the HSE mass.
While V06, G07 and S09 applied temperature deprojection and detailed temperature
profile modeling, E11 did not perform temperature deprojection. The density profile
and the projected temperature profile were modeled with a double $\beta$-model
and a powerlaw respectively. Both models are substantially simpler than the ones
adopted in the other work and some fits in E11 are not good. 

Thus, the results from V06, G07 and S09 agree to $\sim$ 20\% on the gas fraction and the
total mass, and to $\sim$ 10\% on the gas mass. The E11 results are further off, especially
for some E11 groups with very low gas fraction and gas mass.
We suspect that the difference is from the background analysis as most of the gas mass
is from low surface brightness regions.
E11 did not present any detail on their background analysis and
did not publish the detailed fits for individual groups so more in-depth
examination is beyond the scope of this paper.

\end{appendix}

\section*{References}
\begin{harvard}

\item[] Akritas M. G. \& Bershady M. A. 1996 {\it Astrophys. J.} {\bf 470} 706
\item[] Allen S. W., Schmidt R. W., Fabian A. C. 2002 {\it Mon. Not. R. Astron. Soc.} {\bf 334} L11
\item[] Anders E. \& Grevesse N. 1989 {\it Geochimica et Cosmochimica Acta} {\bf 53} 197
\item[] Arnaud M. \& Evrard A. E. 1999 {\it Mon. Not. R. Astron. Soc.} {\bf 305} 631
\item[] Arnaud M., Pointecouteau E., Pratt G. W. 2005 {\it Astron. Astrophys.} {\bf 441} 893
\item[] Arnaud M. {\it et al.} 2010 {\it Astron. Astrophys.} {\bf 517} 92
\item[] Baumgartner W. H., Loewenstein M., Horner D. J., Mushotzky R. F. 2005 {\it Astrophys. J.} {\bf 620} 680
\item[] B\^irzan L. {\it et al.} 2008 {\it Astrophys. J.} {\bf 686} 859
\item[] B\"ohringer H. {\it et al.} 2007 {\it Astron. Astrophys.} {\bf 469} 363	
\item[] Bower R. G., McCarthy I. G., Benson A. J. 2008 {\it Mon. Not. R. Astron. Soc.} {\bf 390} 1399
\item[] Buote D. A. 2000a {\it Mon. Not. R. Astron. Soc.} {\bf 311} 176
\item[] Buote D. A. 2000b {\it Astrophys. J.} {\bf 539} 172
\item[] Buote D. A. {\it et al.} 2004 {\it Astrophys. J.} {\bf 607} L91
\item[] Buote D. A., \& Humphrey P. J.\ 2012 {\it Astrophysics and Space Science Library} {\bf 378} 235
\item[] Burenin R. A. {\it et al.} 2007 {\it Astrophys. J. Supp.} {\bf 172} 561
\item[] Cavagnolo K. W. {\it et al.} 2008 {\it Astrophys. J.} {\bf 683} L107
\item[] Cavagnolo K. W. {\it et al.} 2009 {\it Astrophys. J. Supp.} {\bf 182} 12
\item[] Cavagnolo K. W. {\it et al.} 2010 {\it Astrophys. J.} {\bf 720} 1066
\item[] Chen Y. {\it et al.} 2007 {\it Astron. Astrophys.} {\bf 466} 805
\item[] Crawford C. S. {\it et al.} 1999 {\it Mon. Not. R. Astron. Soc.} {\bf 306} 857
\item[] Croston J. H. {\it et al.} 2005 {\it Mon. Not. R. Astron. Soc.} {\bf 357} 279
\item[] Croston J. H. {\it et al.} 2008 {\it Mon. Not. R. Astron. Soc.} {\bf 386} 1709
\item[] Croston J. H. {\it et al.} 2008 {\it Astron. Astrophys.} {\bf 487} 431
\item[] Dai Xinyu {\it et al.} 2010 {\it Astrophys. J.} {\bf 719} 119
\item[] D\'emocl\`es J. {\it et al.} 2010 {\it Astron. Astrophys.} {\bf 517} 52
\item[] Diehl S., Statler T. S. 2008 {\it Astrophys. J.} {\bf 687} 986
\item[] Dietrich J. P. {\it et al.} 2009 {\it Astron. Astrophys.} {\bf 499} 669
\item[] Dunkley J. {\it et al.} 2011 {\it Astrophys. J.} {\bf 739} 52
\item[] Dunn R. J. H. {\it et al.} 2010 {\it Mon. Not. R. Astron. Soc.} {\bf 404} 180
\item[] Eckert D. {\it et al.} 2011 {\it Astron. Astrophys.} {\bf 526} 79
\item[] Eckert D. {\it et al.} 2011 arXiv:1111.0020
\item[] Eckmiller H. J. {\it et al.} 2011 {\it Astron. Astrophys.} {\bf 535} 105 (E11)
\item[] Ettori S., Balestra I. 2009 {\it Astron. Astrophys.} {\bf 496} 343
\item[] Fabjan D. {\it et al.} 2010 {\it Mon. Not. R. Astron. Soc.} {\bf 401} 1670
\item[] Ferland G. J. {\it et al.} 2009 {\it Mon. Not. R. Astron. Soc.} {\bf 392} 1475
\item[] Finoguenov A. {\it et al.} 2006 {\it Astrophys. J.} {\bf 646} 143
\item[] Finoguenov A. {\it et al.} 2007a {\it Mon. Not. R. Astron. Soc.} {\bf 374} 737
\item[] Finoguenov A. {\it et al.} 2007b {\it Astrophys. J. Supp.} {\bf 172} 182
\item[] Freeland E. \& Wilcots E. 2011 {\it Astrophys. J.} {\bf 738} 145
\item[] Fukazawa Y. {\it et al.} 2006 {\it Astrophys. J.} {\bf 636} 698
\item[] Gaspari M. {\it et al.} 2011 {\it Mon. Not. R. Astron. Soc.} {\bf 415} 1549
\item[] Gastaldello F. {\it et al.} 2007 {\it Astrophys. J.} {\bf 669} 158 (G07)
\item[] Giodini S. {\it et al.} 2009 {\it Astrophys. J.} {\bf 703} 982
\item[] Gitti M. {\it et al.} 2010 {\it Astrophys. J.} {\bf 714} 758
\item[] Gitti M., Brighenti F., McNamara B. R. 2012 {\it Advances in Astronomy} {\bf 2012} 1
\item[] Gonzalez A. H., Zaritsky D., Zabludoff A. I. 2007 {\it Astrophys. J.} {\bf 666} 147
\item[] Helsdon S. F., Ponman T. J. 2000 {\it Mon. Not. R. Astron. Soc.} {\bf 315} 356
\item[] Helsdon S. F., Ponman T. J., Mulchaey J. S. 2005 {\it Astrophys. J.} {\bf 618} 679
\item[] Hudson D. S. {\it et al.} 2010 {\it Astron. Astrophys.} {\bf 513} 37
\item[] Humphrey P. J. {\it et al.} 2006 {\it Astrophys. J.} {\bf 646} 899
\item[] Humphrey P. J. {\it et al.} 2011 {\it Astrophys. J.} {\bf 729} 53
\item[] Humphrey P. J. {\it et al.} 2012 {\it Astrophys. J.} {\bf 748} 11
\item[] Johnson R., Ponman T. J., Finoguenov A. 2009 {\it Mon. Not. R. Astron. Soc.} {\bf 395} 1287
\item[] Johnson R. {\it et al.} 2011 {\it Mon. Not. R. Astron. Soc.} {\bf 413} 2467
\item[] Khosroshahi H. G. {\it et al.} 2004 {\it Mon. Not. R. Astron. Soc.} {\bf 349} 1240
\item[] Koester B. P. {\it et al.} 2007 {\it Astrophys. J.} {\bf 660} 239
\item[] Komatsu E. {\it et al.} 2011 {\it Astrophys. J. Supp.} {\bf 192} 18
\item[] Komiyama M. {\it et al.} 2009 {\it PASJ} {\bf 61} 337
\item[] Leauthaud A. {\it et al.} 2010 {\it Astrophys. J.} {\bf 709} 97
\item[] Leauthaud A. {\it et al.} 2012 {\it Astrophys. J.} {\bf 746} 95
\item[] Leccardi A. \& Molendi S. 2008a {\it Astron. Astrophys.} {\bf 486} 359
\item[] Leccardi A. \& Molendi S. 2008b {\it Astron. Astrophys.} {\bf 487} 461
\item[] Lin Y. T., Mohr J. J. \& Stanford, S. A. 2003 {\it Astrophys. J.} {\bf 591} 749
\item[] Lin Y. T. \& Mohr J. J. 2004 {\it Astrophys. J.} {\bf 617} 879
\item[] von der Linden A. {\it et al.} 2010 {\it Mon. Not. R. Astron. Soc.} {\bf 404} 1231
\item[] Lloyd-Davies E. J. {\it et al.} 2000 {\it Mon. Not. R. Astron. Soc.} {\bf 315} 689
\item[] Lueker M. {\it et al.} 2010 {\it Astrophys. J.} {\bf 719} 1045
\item[] Mahdavi A. {\it et al.} 2005 {\it Astrophys. J.} {\bf 622} 187
\item[] Mathews W. G., Brighenti F. 2003 {\it ARA\&A} {\bf 41} 191
\item[] Maughan B. J. {\it et al.} 2008 {\it Astrophys. J. Supp.} {\bf 174} 117
\item[] McCarthy I. G. {\it et al.} 2010 {\it Mon. Not. R. Astron. Soc.} {\bf 406} 822
\item[] McCarthy I. G. {\it et al.} 2011 {\it Mon. Not. R. Astron. Soc.} {\bf 412} 1965
\item[] McDonald M., Veilleux S., Mushotzky R. 2011 {\it Astrophys. J.} {\bf 731} 33
\item[] McGee S. L., Balogh M. L. 2010 {\it Mon. Not. R. Astron. Soc.} {\bf 403} L79
\item[] McNamara B. R. \& Nulsen P. E. J. 2007 {\it ARA\&A} {\bf 45} 117
\item[] McNamara B. R. {\it et al.} 2012 \NJP, this Focus issue
\item[] Motl P. M. {\it et al.} 2004 {\it Astrophys. J.} {\bf 606} 635
\item[] Mulchaey J. S. 2000, {\it ARA\&A} {\bf 38} 289
\item[] Mulchaey J. S. {\it et al.} 2003 {\it Astrophys. J. Supp.} {\bf 145} 39
\item[] Murakami H. {\it et al.} 2011 {\it PASJ} {\bf 63} 963
\item[] Nagai D., Kravtsov A. V., Vikhlinin A. 2007 {\it Astrophys. J.} {\bf 668} 1
\item[] Nevalainen J., David L., Guainazzi M. 2010 {\it Astron. Astrophys.} {\bf 523} 22
\item[] O'Hara T. B. {\it et al.} 2006 {\it Astrophys. J.} {\bf 639} 64
\item[] Osmond J. P. F. \& Ponman, T. J. 2004 {\it Mon. Not. R. Astron. Soc.} {\bf 350} 1511 (OP04)
\item[] O'Sullivan E. {\it et al.} 2011a {\it Astrophys. J.} {\bf 735} 11
\item[] O'Sullivan E. {\it et al.} 2011b {\it Mon. Not. R. Astron. Soc.} {\bf 416} 2916
\item[] Ponman T. J., Cannon D. B., Navarro J. F. 1999 {\it Nature} {\bf 397} 135
\item[] Ponman T. J. {\it et al.} 2003 {\it Mon. Not. R. Astron. Soc.} {\bf 343} 331
\item[] Popesso P. {\it et al.} 2007 {\it Astron. Astrophys.} {\bf 461} 397
\item[] Pratt G. W. {\it et al.} 2007 {\it Astron. Astrophys.} {\bf 461} 71
\item[] Pratt G. W. {\it et al.} 2009 {\it Astron. Astrophys.} {\bf 498} 361 (P09)
\item[] Pratt G. W. {\it et al.} 2010 {\it Astron. Astrophys.} {\bf 511} 85
\item[] Puchwein E., Sijacki D., Springel V. 2008 {\it Astrophys. J.} {\bf 687} L53
\item[] Rafferty D. A., McNamara B. R., Nulsen P. E. J. 2008 {\it Astrophys. J.} {\bf 687} 899
\item[] Randall, S. W. {\it et al.} 2011 {\it Astrophys. J.} {\bf 726} 86
\item[] Rasia E. {\it et al.} 2008 {\it Astrophys. J.} {\bf 674} 728
\item[] Rasmussen J. {\it et al.} 2006 {\it Mon. Not. R. Astron. Soc.} {\bf 373} 653
\item[] Rasmussen J. \& Ponman, T. J. 2007 {\it Mon. Not. R. Astron. Soc.} {\bf 380} 1554
\item[] Rasmussen J. \& Ponman, T. J. 2009 {\it Mon. Not. R. Astron. Soc.} {\bf 399} 239
\item[] Rasmussen J. {\it et al.} 2010 {\it Astrophys. J.} {\bf 717} 958
\item[] Reichardt C. L. {\it et al.} 2011 arXiv:1111.0932
\item[] Renzini A. 1997 {\it Astrophys. J.} {\bf 488} 35
\item[] Rozo E. {\it et al.} 2009 {\it Astrophys. J.} {\bf 699} 768
\item[] Rykoff E. S. {\it et al.} 2008a {\it Astrophys. J.} {\bf 675} 1106
\item[] Rykoff E. S. {\it et al.} 2008b {\it Mon. Not. R. Astron. Soc.} {\bf 387} L28
\item[] Sanderson A. J. R. {\it et al.} 2003 {\it Mon. Not. R. Astron. Soc.} {\bf 340} 989
\item[] Sanderson A. J. R. {\it et al.} 2006 {\it Mon. Not. R. Astron. Soc.} {\bf 372} 1496
\item[] Sato K. {\it et al.} 2010 {\it PASJ} {\bf 62} 1445
\item[] Sehgal N. {\it et al.} 2010 {\it Astrophys. J.} {\bf 709} 920
\item[] Shaw L. D. {\it et al.} 2010 {\it Astrophys. J.} {\bf 725} 1452
\item[] Shirokoff E. {\it et al.} 2011 {\it Astrophys. J.} {\bf 736} 61
\item[] Sun M {\it et al.} 2003 {\it Astrophys. J.} {\bf 598} 250
\item[] Sun M {\it et al.} 2007 {\it Astrophys. J.} {\bf 657} 197
\item[] Sun M {\it et al.} 2009 {\it Astrophys. J.} {\bf 693} 1142 (S09)
\item[] Sun M 2009 {\it Astrophys. J.} {\bf 704} 1586 (S09)
\item[] Sun M {\it et al.} 2011 {\it Astrophys. J.} {\bf 727} L49
\item[] Tonry J. L. {\it et al.} 2001 {\it Astrophys. J.} {\bf 546} 681
\item[] Trac H., Bode P., Ostriker J. P. 2011 {\it Astrophys. J.} {\bf 727} 94
\item[] Vikhlinin A. {\it et al.} 1999 {\it Astrophys. J.} {\bf 525} 47
\item[] Vikhlinin A. {\it et al.} 2006 {\it Astrophys. J.} {\bf 640} 691 (V06)
\item[] Vikhlinin A. {\it et al.} 2009 {\it Astrophys. J.} {\bf 692} 1033 (V09)
\item[] Voit G. M. 2005 {\it Rev. Mod. Phys.}, {\bf 77} 207
\item[] Voit G. M., Kay S. T., Bryan G. L. 2005 {\it Mon. Not. R. Astron. Soc.} {\bf 364} 909
\item[] Wetzel A. R., Tinker J. L., Conroy C. 2011 arXiv:1107.5311

\end{harvard}

\end{document}